\documentclass[traditabstract]{aa}
\usepackage{graphicx}
\usepackage{natbib}
\newcommand{\kms}{\mbox{km~s$^{-1}$}}
\newcommand{\mols}{\mbox{molec.~s$^{-1}$~}}
\renewcommand{\deg}{\mbox{$^{\circ}$}}

\begin{document}

\title{Molecular investigations of comets C/2002~X5 (Kudo-Fujikawa),
C/2002~V1 (NEAT), and C/2006 P1 (McNaught) at small heliocentric distances}

\author{N. Biver\inst{1} \and D. Bockel\'ee-Morvan\inst{1}
    \and P. Colom\inst{1} \and J. Crovisier\inst{1}
    \and G. Paubert\inst{2}
    \and A. Weiss\inst{3}
    \and H. Wiesemeyer\inst{3}
}

   \offprints{N. Biver}
\institute{LESIA, CNRS UMR 8109, UPMC, Universit\'e Paris-Diderot, Observatoire de Paris, 5 pl. Jules Janssen, F-92195 Meudon, France  \email{Nicolas.Biver@obspm.fr}
\and IRAM, Avd. Divina Pastora, 7, 18012 Granada, Spain
\and Max Planck Institute for Radio Astronomy, P.O. Box 20 24, 53010 Bonn, Germany}

\date{\today}

\abstract{We present unique spectroscopic radio observations of
comets C/2002~X5 (Kudo-Fujikawa), C/2002~V1 (NEAT), and 
C/2006~P1 (McNaught), which came within $r_h \approx$ 0.2 AU
of the Sun in 2003 and 2007.
The molecules OH, HCN, HNC, CS, and CH$_3$OH were detected in 
each of these comets when they were exposed to strong
heating from the Sun. Both HC$_3$N and HCO$^+$ were detected 
in comets C/2002~X5 and C/2006 P1, respectively.
We show that in these very productive comets close to the Sun 
screening of the photodissociation by the Sun UV radiation
plays a non-negligible role. Acceleration of the gas expansion 
velocity and day-night asymmetry is also measured and modeled.
The CS photodissociation lifetime was constrained to be about 
$2.5\times10^{-5}$~s$^{-1}$ at $r_h=1$ AU. The relative abundances are
compared to values determined from more distant observations of
C/2002~X5 or other comets. A high HNC/HCN production-rate ratio,
in the range 10--30\% between 0.5 and 0.1 AU from the Sun, is
measured. The trend for a significant enrichment in CS in cometary
comae (CS/HCN$\propto r_h^{-0.8}$) is confirmed in all three
comets. The CH$_3$OH/HCN production rate ratio decreases at low
$r_h$.  The HC$_3$N/HCN production rate ratio in comet C/2002~X5
is four times higher than measured in any other comet.
}

\authorrunning{Biver et al.}

\titlerunning{Comets C/2002~X5, C/2002~V1, and C/2006~P1 close to the Sun}
\maketitle


\section{Introduction}
The composition of cometary nuclei is of strong interest in
understanding their origin. Having spent most of their time in a
very cold environment, these objects should not have evolved much
since their formation. Thus, their composition provides clues to
the composition in the outer regions of the Solar Nebula where
they formed. The last two decades have proven the efficiency of
microwave spectroscopy in investigating the chemical composition
of cometary atmospheres. About 20 different cometary molecules
have now been identified at radio wavelengths \citep{Boc04a}.

In this paper, we extend our investigations of the composition of
cometary atmospheres from radio observations 
\citep{Boc04a,Biv02a,Biv06a,Biv07b} to three comets
observed in 2003 and 2007. These observations were designed to
measure the molecular abundances of comets approaching close to the
Sun to investigate how the strength of the solar heating 
of both the comet nucleus and of its environment affects the coma 
composition. 
Previous observations have suggested that the relative production 
rates of several molecules vary with heliocentric distance
\citep{Biv06a}. The passages of comets C/2002~X5 (Kudo-Fujikawa),
C/2002~V1 (NEAT), and C/2006~P1 (McNaught) provided us with the
opportunity to measure molecular abundances at heliocentric
distances ($r_h$) between 0.1 and 0.25 AU, about one order of
magnitude smaller than usual. This study is complementary to the
long-term monitoring of comet C/1995~O1 (Hale-Bopp) \citep{Biv02b},
which provided information on the outgassing of a comet between 0.9
and 14 AU.

Opportunities to plan observations of comets passing within 0.2 AU
from the Sun are rare. For example, comet C/1998~J1 (SOHO) was
discovered too late to establish an accurate ephemeris around
perihelion time. In addition, Sun-grazing comets often do not
survive and even disintegrate before reaching perihelion, making
observing plans extremely difficult. The last opportunity was comet
C/1975~V1 (West). The observations reported here are unique, and
required to develop specific observing strategies and analyses.

The organization of the paper is as follows. In Sect. 2, we
present the observations of comets C/2002~X5 (Kudo-Fujikawa),
C/2002~V1 (NEAT), and C/2006~P1 (McNaught) performed with the 30-m
telescope of the Institut de Radioastronomie Millim\'etrique
(IRAM) and the Nan\c{c}ay radio telescope. In Sects. 3 and 4, the
analysis of these observations is presented. A summary follows in
Sect. 5.


\section{Observations}
    Owing to the late discovery or assessment of their activity, the three
comets were observed as targets of opportunity using the IRAM 30-m
and Nan\c{c}ay radio telescopes. This was possible because these
telescopes do not have tight solar elongation constraints. The
small solar elongation ($<$ 10\degr~at $r_h$ $<$ 0.2 AU) resulted
in limited observing support from other observatories.

\begin{table*}
\caption[]{Observing conditions and ephemeris errors}\label{tabefe}
\begin{center}
\begin{tabular}{llccccccc}
\hline
Comet & UT date     & $r_{h}$ & $\Delta$ & Phase
    & \multicolumn{2}{c}{$O-C_u$$^{\footnotesize{1}}$ }
    & \multicolumn{2}{c}{$C_f-C_u$$^{\footnotesize{1}}$} \\

      & [yyyy/mm/dd.d]  & [AU]   & [AU] & [\deg]
    & $\delta$RA & $\delta$Dec. & $\delta$RA & $\delta$Dec. \\
\hline
C/2002~X5 & 2003/01/13.6 & 0.553 & 1.032 & 69.3
    & - & - & +1.6\arcsec & --2.2\arcsec \\
          & 2003/01/26.5 & 0.214 & 1.172 & 26.1
    & 1.5\arcsec & --11\arcsec & +0.9\arcsec & --10.3\arcsec \\
          & 2003/03/12.7 & 1.184 & 1.096 & 50.5
    & - & - & +3.0\arcsec & +0.9\arcsec \\
C/2002~V1 & 2003/02/16.6 & 0.135 & 0.978 & 90.3
    & --1.9\arcsec & --3.5\arcsec & --3.9\arcsec & --6.6\arcsec \\
          & 2003/02/17.6 & 0.107 & 0.986 & 87.7
    & --1.5\arcsec & --4.5\arcsec & --2.5\arcsec & --8.8\arcsec \\
C/2006~P1 & 2007/01/15.6 & 0.207 & 0.817 & 140.0
    & +1\arcsec & +30\arcsec & --1.2\arcsec & +27.1\arcsec \\
      & 2007/01/16.6 & 0.229 & 0.822 & 129.3
    & +1\arcsec & +28\arcsec & +0.6\arcsec & +25.7\arcsec \\
      & 2007/01/17.6 & 0.256 & 0.833 & 119.1
    & +0\arcsec & +23\arcsec & +2.0\arcsec & +23.5\arcsec \\
\hline
\end{tabular}
\end{center}
$^1$ $O$, $C_u$, and $C_f$ denote the position of the HCN peak
emission determined from mapping, the used computed ephemeris, and
the final (reference) ephemeris, respectively. Orbital elements
are given for
C/2002~X5 (MPEC 2003-A41, 2003-A84, and 2003-D20 used, JPL\#40$^2$
reference), C/2002~V1 (MPEC 2003-C42 used, JPL\#34 reference), C/2006~P1
(JPL\#15, JPL\#25 reference). The uncertainty in reference orbit
is 2\arcsec.\\
$^2$: JPL HORIZONS ephemerides: http://ssd.jpl.nasa.gov/?ephemerides
\end{table*}

\subsection{C/2002~X5 (Kudo-Fujikawa)}
Comet C/2002~X5 (Kudo-Fujikawa) was discovered visually at $m_v$ =
9 on 13--14 December 2002 by two Japanese amateur astronomers, T.
Kudo and S. Fujikawa \citep{iauc8032}. At about 1.2 AU from the
Earth and the Sun at that time, it was then a moderately active
comet. It passed perihelion on 29 January 2003 at a perihelion
distance $q$ = 0.190 AU from the Sun.

\begin{figure}
\centering
\resizebox{\hsize}{!}{\includegraphics[angle=0]{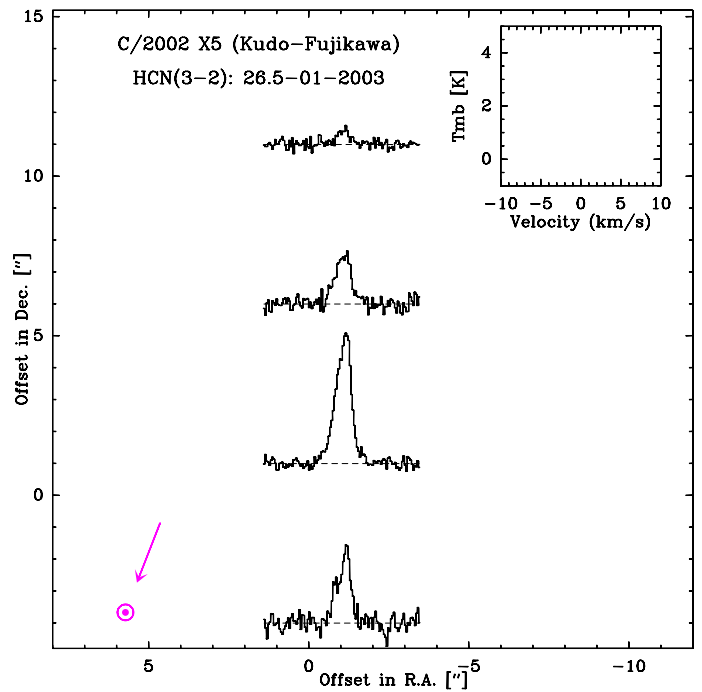}}
\caption{Coarse map of the HCN $J =$ 3--2 emission from comet C/2002~X5
(Kudo-Fujikawa) observed with the IRAM 30-m. Positional offsets in arcsec are 
with respect to the final ephemeris (Table~\ref{tabefe}). 
All four spectra are plotted at the same intensity and velocity scales, 
provided in the upper right box. The projected direction of the Sun
is also given.} \label{spec02x5map}
\end{figure}

\begin{figure}
\centering
\resizebox{\hsize}{!}{\includegraphics[angle=270]{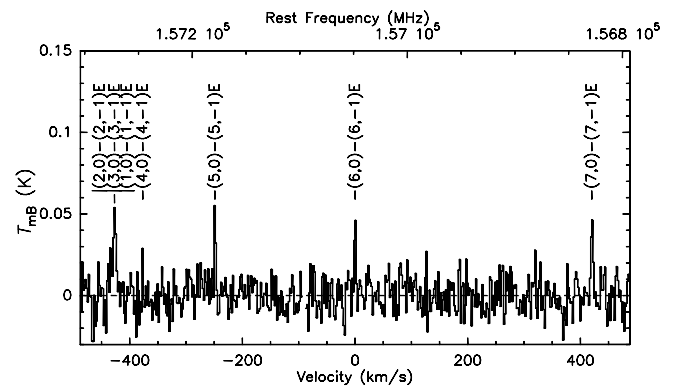}}
\caption{CH$_3$OH lines at 157~GHz observed with the IRAM 30-m
telescope in comet C/2002~X5 (Kudo-Fujikawa) on 26.5 January
2003. The vertical scale is main-beam brightness temperature. The
horizontal scales are the rest frequency (upper scale) or the
Doppler velocity in the comet rest-frame, relative to the
($6_0-6_{-1}$)E line frequency (bottom scale).}
\label{spec02x5-ch3oh}
\end{figure}

\begin{figure}
\centering
\resizebox{\hsize}{!}{\includegraphics[angle=270]{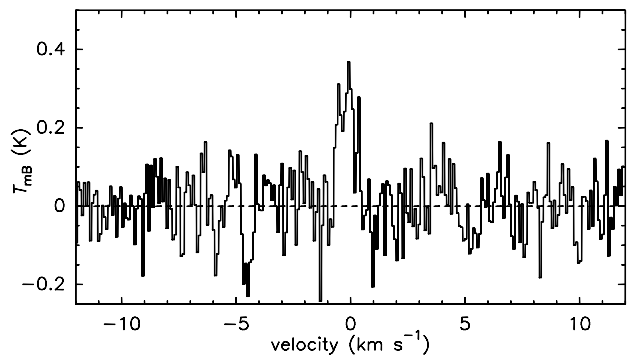}}
\caption{HCN $J =$ 3--2 line observed with the IRAM 30-m telescope
in comet C/2002~X5 (Kudo-Fujikawa) on 12.7 March 2003. The
vertical scale is the main beam brightness temperature. The
horizontal scale is the Doppler velocity in the comet rest-frame.}
\label{spec02x5-hcnmar}
\end{figure}

\begin{figure}
\centering
\resizebox{\hsize}{!}{\includegraphics[angle=0]{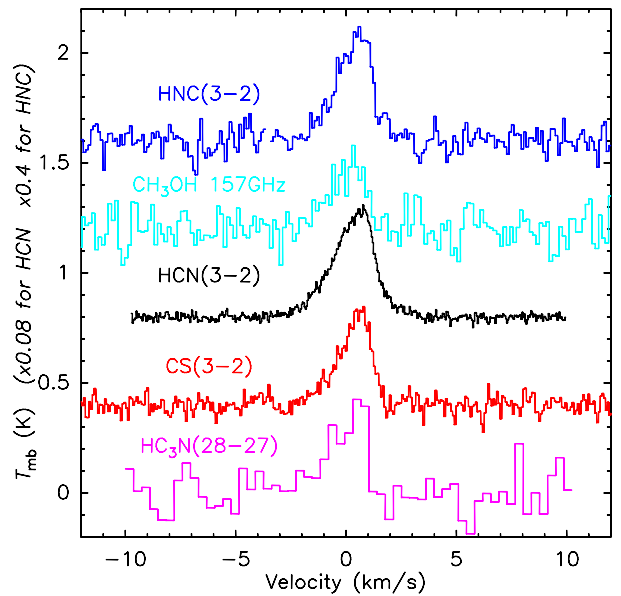}}
\caption{HC$_3$N, CS, HCN, CH$_3$OH (sum of the 3 brightest lines
at 157~GHz) and HNC lines observed with the IRAM 30-m telescope in
comet C/2002~X5 (Kudo-Fujikawa) on 26.5 January 2003. The spectra
are plotted with the same horizontal scale to compare the line
widths. The intensities of the spectra of HCN and HNC were divided 
by 12.5 and 2.5, respectively. Spectra are shifted vertically for clearer 
viewing.}
\label{spec02x5-5lines}
\end{figure}

One of the main goals of the observations of comet C/2002~X5 at the
IRAM 30-m was to measure the evolution of the HNC/HCN production
rate ratio as it approached the Sun. A first observing slot was
scheduled on 4--5 January 2003, at $r_h$ = 0.8 AU, but the weather
prevented any observation. Observations performed on 13 January
($r_h=0.55$ AU) only partly succeeded because of technical problems.
Most data were acquired on 26.5 January, 2 days before perihelion, at
$r_h$ = 0.21 AU. At that time, the comet was only observable
visually by the C3 coronagraph aboard the SOlar Heliospheric
Observatory (SOHO) --- the solar elongation was 5.5\degr~
\citep{Bou03,Pov03}. The pointing of the comet was a
real challenge, since the ephemeris uncertainty was expected to be
on the order of the beam size (10--20\arcsec).  
Since at $r_h$ = 0.2 AU molecular lifetimes are
typically shorter than an hour, hence photodissociation scale
lengths are smaller than the beam size, accurate pointing
was required. From coarse mapping, we
found the comet about 10\arcsec~south of its predicted position
(Fig.~\ref{spec02x5map}, Table~\ref{tabefe}). 
The last observing run at IRAM took
place on 12 March 2003 ($r_h$ = 1.2 AU,
Fig.~\ref{spec02x5-hcnmar}), when the comet was receding from the
Sun and about 100 times less productive. A log of the observations
and measured line areas are given in Table~\ref{tabobs02x5}.
Sample spectra are shown in
Figs~\ref{spec02x5-ch3oh}--\ref{spec02x5-5lines}.

To monitor the water production rate, observations of OH
at 18-cm using the Nan\c{c}ay radio telescope were scheduled on a
daily basis from 1 January to 10 April 2003. The OH lines were only
detected when the comet was between 1 and 0.4 AU from the Sun,
inbound and outbound.

\subsection{C/2002~V1 (NEAT)}

Comet C/2002~V1 (NEAT) was discovered on 6 November 2002 by the
Near Earth Asteroid Tracking (NEAT) program telescope on the
Haleakala summit of Maui, Hawaii \citep{iauc8010}. At that time,
it was a relatively faint object ($m_v$ = 17).  Comet C/2002~V1
(NEAT) brightened rapidly after its discovery, becoming a
potentially interesting target for observations at perihelion on
18 February 2003 at $q$ = 0.099 AU. The orbital period was 
estimated to be 9000 years (Nakano note NK965
\footnote{http://www.oaa.gr.jp/~oaacs/nk/nk965.htm}), 
suggesting that it has survived a close
passage to the Sun at its previous perihelion and
might be expected to do so again.

\begin{figure}
\centering
\resizebox{\hsize}{!}{\includegraphics[angle=0]{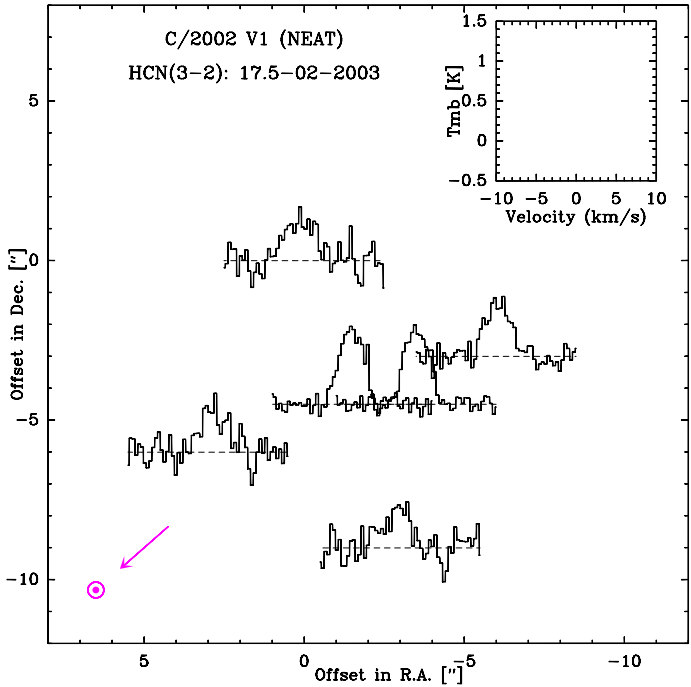}}
\caption{Scales as in Fig.~\ref{spec02x5map}: coarse map of the 
HCN $J$ = 3--2 emission in comet C/2002~V1 (NEAT) with the 
IRAM 30-m on 17.5 February 2003. But in this case, spectra
are plotted at positions relative to the ephemerides used
during the observations.}
\label{spec02v1map}
\end{figure}

The observations at IRAM were undertaken on 16 and 17 February 2003
($r_h$ = 0.13--0.11 AU), i.e., just one day before perihelion. As
for comet C/2002~X5 (Kudo-Fujikawa), the observations were
challenging because of the lack of supporting optical astrometry, the
comet being 8--6$\deg$ away from the Sun and only seen by the SOHO
coronagraph when it became as bright as magnitudes $-1$ to $-2$.
A coarse map of the HCN $J$=3--2 line at IRAM revealed the comet
to be at about 5.5\arcsec~(half a beam, Fig.~\ref{spec02v1map}, 
Table~\ref{tabefe})
from its expected position. Observational data are given in
Table~\ref{tabobs02v1}.

OH 18-cm observations were obtained nearly every day from 31 December
2002 to 25 March 2003. The comet was only detected at 
$r_h > 0.4$ AU from the Sun, though the water production rate
($Q_{\rm H_2O}$) likely exceeded $2\times10^{30}$ \mols 
at perihelion (Sect.3).

\begin{figure}
\centering
\resizebox{\hsize}{!}{\includegraphics[angle=0]{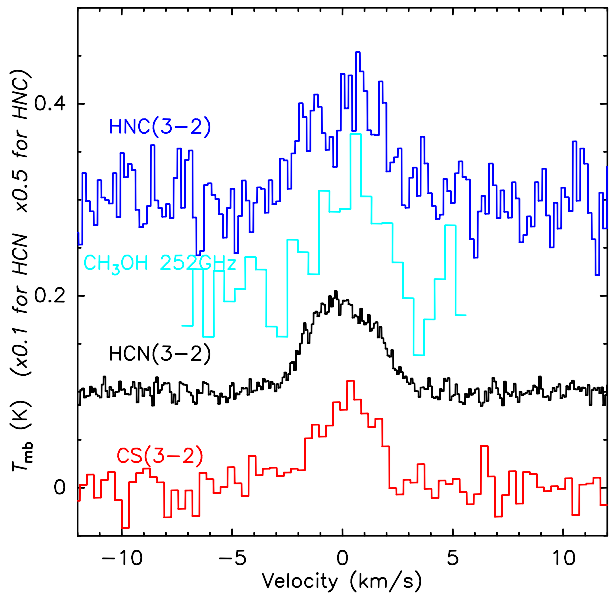}}
\caption{CS, HCN, CH$_3$OH (sum of the 8 brightest lines at
252~GHz), and HNC lines observed with the IRAM 30-m telescope in
comet C/2002~V1 (NEAT) on 16.5--17.6 February 2003. The spectra
are plotted with the same horizontal scale to compare the line
widths. The intensities of the spectra of HCN and HNC were divided 
by 10 and 2, respectively. Spectra are shifted vertically for clearer viewing.}
\label{spec02v1-4lines}
\end{figure}

\subsection{C/2006~P1 (McNaught)}
C/2006~P1 was discovered by Robert McNaught at $m_v$ = 17.3 on 7
August 2006 \citep{iauc8737} as it was at 3.1 AU from the Sun.
The geometry was very unfavorable for observing this comet as it
approached the Sun. The comet was basically lost in the glare of
the Sun in November and December 2006 from $r_h$ = 1.5 to 0.5 AU.
At $r_h$ $>$ 0.5 AU, its intrinsic brightness was similar to that
of comet C/1996 B2 (Hyakutake). It brightened rapidly in early
January 2007 to peak at $m_v$ $\sim$ --5 and became visible to the
naked eye in broad daylight \citep{iauc8796}. It passed
perihelion on 12 January 2007 at $q$ = 0.17 AU. Comet  C/2006~P1
(McNaught) was the brightest and most productive comet since
C/1965 S1 (Ikeya-Seki). The absence of an ion tail led
\citet{Ful07} to argue that, because of a very high outgassing rate,
the diamagnetic cavity was so large that ions were
photodissociated before reaching the region where they could
interact with the solar wind. A week after perihelion, C/2006~P1
displayed a fantastic dust tail with many striae curving around
1/3 of sky at a mean distance of 20\degr~from the Sun.

\begin{figure}
\centering
\resizebox{\hsize}{!}{\includegraphics[angle=0]{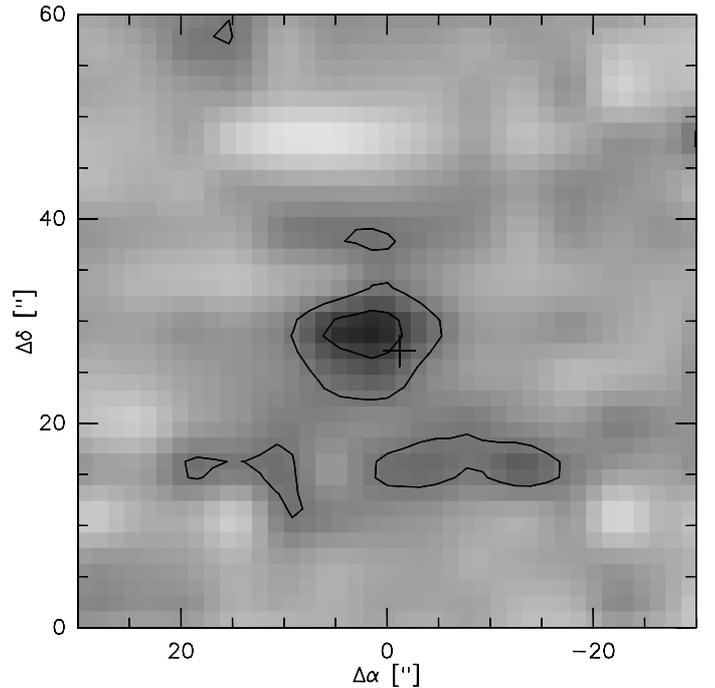}}
\caption{On-the-fly map of the HCN $J$ = 3--2 emission in comet
C/2006~P1 (McNaught) obtained with the IRAM 30-m on 15.65 January
2007 (4 min. integration time). The de-pixelized image shows the line 
area between $-3$ and +3 \kms, contour levels are 2 and 4$\sigma_{\rm rms}$. 
The cross marks 
the comet position as computed from the reference ephemeris.
Before gridding to beam/2 sampling, a linear spatial baseline 
was subtracted from each on-the-fly subscan to remove total power 
fluctuations of mainly atmospheric orgin.} \label{img06p1hcn}
\end{figure}

\begin{figure}
\centering
\resizebox{\hsize}{!}{\includegraphics[angle=0]{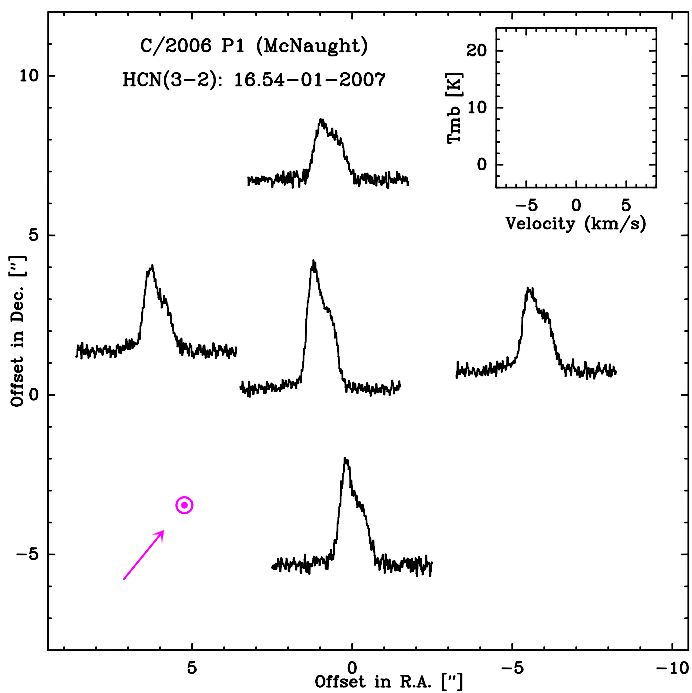}}
\caption{Coarse map of the HCN $J$(3--2) emission from comet
C/2006~P1 (McNaught) on 16.5 January 2007, displayed as in 
Figs~\ref{spec02x5map} and \ref{spec02v1map}, with positions relative
to the final ephemerides (Table~\ref{tabefe}).}
\label{spec06p1map}
\end{figure}

The IRAM observations were performed on 15, 16, and 17
January 2007 ($r_h$ = 0.21--0.25 AU).  The comet ephemeris was
expected to be possibly wrong by up to 1\arcmin~(corresponding to
the 3-$\sigma$ uncertainty from the JPL's HORIZONS ephemeris 
\citep{horizons}). 
An on-the-fly map of the HCN $J$(3--2) line  
on 15.6 January UT with the IRAM 30-m HERA array of
receivers showed that the comet was 30\arcsec~north of its
predicted position (Fig.~\ref{img06p1hcn}). Orbit updates
later confirmed the observed offset (Table~\ref{tabefe}). 
Most observations
consisted of five-point integrations at 0 and 6\arcsec~ offsets
in RA and Dec to pinpoint the maximum emission (e.g.,
Fig.~\ref{spec06p1map}). On 17.5 January, the observations were
affected by strong anomalous refraction due to the low elevation
 of the comet (21--17\deg). The effect was estimated to be
equivalent to a mean pointing offset of up to 12\arcsec, implying
a loss of 90\% of the signal. Observational data are summarized in
Table~\ref{tabobs06p1}.

OH 18-cm observations were performed daily between 10 and 20
January. The comet was detected intermittently. OH 18-cm lines
had never been detected that close to the Sun (0.17 AU) in a comet
before.

\begin{figure}
\centering
\resizebox{\hsize}{!}{\includegraphics[angle=270]{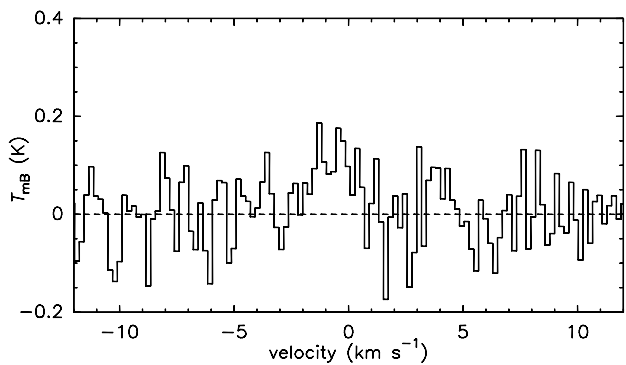}}
\caption{HDO line at 225.896~GHz observed with the IRAM 30-m
telescope in comet C/2006~P1 on 16.6 January 2007. The
vertical scale is the main-beam brightness temperature. The
horizontal scale is the Doppler velocity in the comet rest-frame.} \label{spechdo06p1}
\end{figure}

\begin{figure}
\centering
\resizebox{\hsize}{!}{\includegraphics[angle=0]{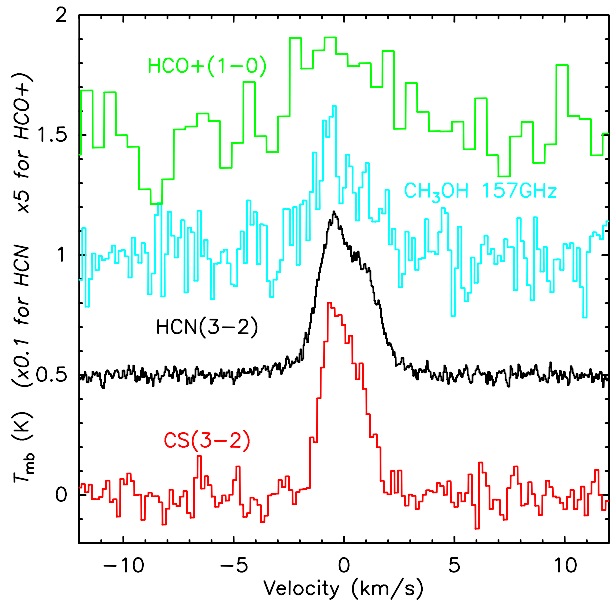}}
\caption{CS, HCN, CH$_3$OH (sum of the 4 brightest lines at
157~GHz), and HNC lines observed with the IRAM 30-m telescope in
comet C/2006~P1 (McNaught) on 16.5 or 17.6 January 2007. Spectra
have been plotted at the same horizontal scale to compare the line
widths. The intensity scale is reduced by 10 and 2 for HCN and HNC, 
respectively, and shifted vertically for
better viewing.} \label{spec06p1-4lines}
\end{figure}


\section{Analysis of Nan\c{c}ay data - H$_2$O production rates}

The characteristics of the Nan\c{c}ay radio telescope and the OH
observations of comets may be found in \citet{Cro02}.
The OH 18-cm lines usually observed in comets are maser lines pumped
by UV solar radiation \citep{Des81}. However, for high water-production 
rates, the maser emission is quenched by collisions in
a large part of the coma, and most of the signal may originate from
thermal emission. The radius of the region where the maser emission
is quenched is estimated to be $50000 r_h\sqrt{\rm Q_{29}}$~km, where
$Q_{29}$ is the water production rate in units of $10^{29}$~\mols
\citep{Ger98}. For production rates above $6\times10^{29}$~\mols and the
heliocentric distances where the comets were observed ($r_h<0.3$ AU), 
most OH radicals dissociate within the quenching zone. 
Hence, the signal is dominated by OH thermal emission. The OH production
rates or upper limits are given in Table~\ref{tabqh2o} and sample spectra 
are provided in Fig.~\ref{fignancay}.

Table~\ref{tabqh2o} also includes water-production rate measurements
from other instruments for comet C/2002~X5. The H$_2$O line at
557~GHz was observed in this comet with Odin during March 2003
\citep{Biv07a}. Observations were conducted with the SOHO
coronagraph spectrometer on the perihelion date: from H
Lyman $\alpha$ measurements, the peak outgassing rate of water is
estimated to $\sim 3\times10^{30}$ \mols \citep{Com08}. According to
\citet{Pov03} and \citet{Bou03}, this comet displayed a strongly
variable activity around perihelion with a $\sim$two day period.
During the perihelion period, the comet was not or only very marginally
detected at Nan\c{c}ay. Averaging the data obtained during the 
five days around perihelion, we obtain a possible 4-$\sigma$ detection 
(subject to baseline uncertainties) suggesting a production rate of
 $\sim 6\times10^{30}$ \mols (Table~\ref{tabqh2o}). 
This is higher than the SOHO estimate, 
but still within the same order of magnitude.  
Power laws fitted to Nan\c{c}ay and Odin data (Fig.~\ref{figqp02x5})
yield values in the same range 1.5--5.4$\times10^{30}$ \mols at 0.214 AU
from the Sun. A mean value of 3.5$\times10^{30}$ \mols is adopted.

For C/2002~V1 (NEAT), the pre-perihelion observations can be fitted
by a power law ($0.33\pm0.03\times10^{29}\times r_h^{-1.95\pm0.23}$), 
which extrapolates to $Q_{\rm H_2O} = 1.6 \times10^{30}$ and 
$2.5\times10^{30}$ \mols for 16 and
17 January 2003, respectively, consistent with the upper
limit determined for these days. Extrapolation of SOHO-SWAN 
measurements yield values that are four times higher but indicative 
of abundances relative to water that are abnormaly low for all molecules. 
We assume that $Q_{\rm H_2O}$ = $2.0$ and $2.5\times10^{30}$ \mols, 
respectively, these values being more compatible with the day-to-day 
variations observed in millimeter spectra of other molecules.

As for C/2006~P1 (McNaught), the water production rates for
13 and 19 January deduced from the Nan\c{c}ay data 
agree with the estimated HDO production rate from
the marginally detected line at 225.9~GHz (Fig.~\ref{spechdo06p1}) 
and the HDO/H$_2$O$=6\times10^{-4}$ ratio measured in comets. 
The total production rate varied from $\approx 40\times10^{30}$ 
down to $\approx 5\times10^{30}$~\mols at that time (Sect.3). 

\onltab{2}{
\begin{table*}
\caption[]{Molecular observations in comet C/2002~X5 (Kudo-Fujikawa)}\label{tabobs02x5}
\begin{center}
\begin{tabular}{cccrlccr}
\hline
UT date (2003) & $<r_{h}>$ & $<\Delta>$ & Int. time & Line & $\int T_bdv$    & Velocity offset & offset \\[0cm]
[mm/dd.dd--dd.dd] & [AU] & [AU]  &    [min]  &      & [K~km~s$^{-1}$] & [km~s$^{-1}$]   &        \\
\hline
\multicolumn{8}{l}{\it IRAM~30-m:}\\
\hline
01/13.55--13.66 & 0.553 & 1.032 &  60  & HCN(1-0) & $0.119\pm0.024$ & $-0.15\pm0.25$ & 3.5\arcsec \\
01/13.55--13.66 & 0.553 & 1.032 &  40  & HCN(3-2) & $1.405\pm0.058$ & $+0.02\pm0.08$ & 3.5\arcsec \\
01/26.46--26.46 & 0.214 & 1.172 &  10  & HCN(3-2) & $1.075\pm0.190$ & $+0.22\pm0.22$ & 10\arcsec \\
01/26.47--26.49 & 0.214 & 1.172 &  10  & HCN(3-2) & $4.640\pm0.190$ & $+0.24\pm0.05$ &  6\arcsec \\
01/26.50--26.50 & 0.214 & 1.172 &   5  & HCN(3-2) & $6.046\pm0.266$ & $+0.29\pm0.09$ & 4.5\arcsec \\
01/26.51--26.57 & 0.213 & 1.172 &  90  & HCN(3-2) &$14.268\pm0.084$ & $+0.24\pm0.01$ & 1.0\arcsec \\
01/26.58--26.58 & 0.213 & 1.172 &   5  & HCN(3-2) &$11.940\pm0.335$ & $+0.30\pm0.04$ & 4.0\arcsec \\
03/12.65--12.79 & 1.184 & 1.096 &  90  & HCN(3-2) & $0.275\pm0.027$ & $-0.16\pm0.04$ & 3.0\arcsec \\

01/13.55--13.66 & 0.553 & 1.032 &  20  & HNC(3-2) & $0.326\pm0.101$ & $+0.36\pm0.40$ & 3.5\arcsec \\
01/26.46--26.46 & 0.214 & 1.172 &  10  & HNC(3-2) & $<0.501$        &                & 13\arcsec \\
01/26.47--26.49 & 0.214 & 1.172 &  10  & HNC(3-2) & $0.660\pm0.140$ & $+0.40\pm0.26$ & 6.8\arcsec \\
01/26.50--26.57 & 0.213 & 1.172 &  90  & HNC(3-2) & $2.415\pm0.093$ & $+0.33\pm0.15$ & 2.8\arcsec \\
01/26.58--26.58 & 0.213 & 1.172 &   5  & HNC(3-2) & $1.494\pm0.399$ & $+0.02\pm0.39$ & 4.4\arcsec \\
03/12.65--12.79 & 1.184 & 1.096 &  90  & HNC(3-2) & $<0.106$        &                & 3.0\arcsec \\

01/26.47--26.63 & 0.213 & 1.172 & 135 & CH$_3$CN(8,0-7,0) & $0.068\pm0.016$ & $+0.29\pm0.25$ & 1.6\arcsec \\
                &       &       &    & CH$_3$CN(8,1-7,1) & $0.041\pm0.015$ & $-0.08\pm0.33$ & \\
                &       &       &    & CH$_3$CN(8,2-7,2) & $0.028\pm0.019$ &    &  \\
                &       &       &    & CH$_3$CN(8,3-7,3) & $0.091\pm0.019$ & $-0.03\pm0.27$   &  \\
03/12.65--12.79 & 1.184 & 1.096 &  90  & CH$_3$CN(8-7)$^1$ &  $<0.054$     &         & 3.0\arcsec \\

01/26.60--26.63 & 0.212 & 1.172 &  25  & HC$_3$N(28-27) & $0.700\pm0.112$ & $-0.03\pm0.16$ & 1.5\arcsec \\

01/13.55--13.66 & 0.553 & 1.032 &  60  &  CO(2-1) & $<0.094^1$ &  & 3.5\arcsec \\

01/26.46--26.46 & 0.214 & 1.172 &  10  &  CS(3-2) & $0.278\pm0.052$ & $+0.45\pm0.24$ & 10\arcsec \\
01/26.47--26.49 & 0.214 & 1.172 &  10  &  CS(3-2) & $0.407\pm0.057$ & $+0.23\pm0.18$ & 5.5\arcsec \\
01/26.51--26.51 & 0.214 & 1.172 &   5  &  CS(3-2) & $0.463\pm0.048$ & $+0.04\pm0.13$ & 1.5\arcsec \\
01/26.52--26.57 & 0.213 & 1.172 &  85  &  CS(3-2) & $0.744\pm0.030$ & $+0.31\pm0.05$ & 1.3\arcsec \\
01/26.50--26.58 & 0.213 & 1.172 &  10  &  CS(3-2) & $0.787\pm0.058$ & $+0.10\pm0.09$ & 3.7\arcsec \\
01/26.60--26.63 & 0.212 & 1.172 &  25  &  CS(3-2) & $0.875\pm0.034$ & $+0.25\pm0.05$ & 1.5\arcsec \\
03/12.65--12.79 & 1.184 & 1.096 & 135  &  CS(3-2) & $<0.032$        &                & 3.0\arcsec \\

01/26.60--26.63 & 0.212 & 1.172 &  25  & SiO(6-5) & $<0.516$        &                & 1.5\arcsec \\

01/13.55--13.66 & 0.553 & 1.032 & 60 & CH$_3$OH(1,0-1,-1)E & $0.026\pm0.013$ & $-0.51\pm0.36$ & 3.5\arcsec\\
                &       &       &    & CH$_3$OH(2,0-2,-1)E & $0.065\pm0.018$ & $+0.58\pm0.39$ & \\
                &       &       &    & CH$_3$OH(3,0-3,-1)E & $0.099\pm0.018$ & $-0.33\pm0.21$ & \\
                &       &       &    & CH$_3$OH(4,0-4,-1)E & $0.049\pm0.018$ & $-0.25\pm0.54$ & \\
                &       &       &    & CH$_3$OH(5,0-5,-1)E & $0.065\pm0.018$ & $+0.03\pm0.34$ & \\
                &       &       &    & CH$_3$OH(6,0-6,-1)E & $0.047\pm0.018$ & $+0.25\pm0.38$ & \\
                &       &       &    & CH$_3$OH(7,0-7,-1)E & $0.031\pm0.018$ & $+0.46\pm0.64$ & \\
01/26.47--26.63 & 0.213 & 1.172 & 135& CH$_3$OH(1,0-1,-1)E & $0.072\pm0.027$ & $-0.16\pm0.38$ & 3.0\arcsec\\
                &       &       &    & CH$_3$OH(2,0-2,-1)E & $0.101\pm0.030$ & $+0.75\pm0.41$ & \\
                &       &       &    & CH$_3$OH(3,0-3,-1)E & $0.142\pm0.028$ & $-0.28\pm0.22$ & \\
                &       &       &    & CH$_3$OH(4,0-4,-1)E & $0.113\pm0.030$ & $-0.11\pm0.31$ & \\
                &       &       &    & CH$_3$OH(5,0-5,-1)E & $0.197\pm0.028$ & $+0.30\pm0.17$ & \\
                &       &       &    & CH$_3$OH(6,0-6,-1)E & $0.197\pm0.029$ & $+0.13\pm0.17$ & \\
                &       &       &    & CH$_3$OH(7,0-7,-1)E & $0.214\pm0.028$ & $-0.10\pm0.15$ & \\
03/12.65--12.79 & 1.184 & 1.096 & 135& CH$_3$OH(1,0-1,-1)E & $0.042\pm0.012$ & $+0.24\pm0.12$ & 3.0\arcsec\\
                &       &       &    & CH$_3$OH(2,0-2,-1)E & $0.024\pm0.014$ &  & \\
                &       &       &    & CH$_3$OH(3,0-3,-1)E & $0.011\pm0.013$ &  & \\
                &       &       &    & CH$_3$OH(4,0-4,-1)E & $0.015\pm0.014$ &  & \\
                &       &       &    & CH$_3$OH(5,0-5,-1)E & $0.043\pm0.016$ & $-0.12\pm0.15$ & \\
                &       &       &    & CH$_3$OH(6,0-6,-1)E & $0.036\pm0.012$ & $-0.03\pm0.12$ & \\
                &       &       &    & CH$_3$OH(7,0-7,-1)E & $0.027\pm0.012$ & $-0.03\pm0.16$ & \\
        &   &   &   & sum of 7 lines & $0.150\pm0.037$ & $-0.01\pm0.09$ & \\
\hline
\end{tabular}
\end{center}
$^1$ Sum of CH$_3$CN(8,0-7,0), CH$_3$CN(8,1-7,1) and CH$_3$CN(8,2-7,2) lines.
\end{table*}
}

\onltab{3}{
\begin{table*}
\caption[]{Molecular observations in comet C/2002~V1 (NEAT)}\label{tabobs02v1}
\begin{center}
\begin{tabular}{cccrlccr}
\hline
UT date (2003) & $<r_{h}>$ & $<\Delta>$ & Int. time & Line & $\int T_bdv$    & Velocity offset & offset \\[0cm]
[mm/dd.dd--dd.dd] & [AU] & [AU]  &    [min]  &      & [K~km~s$^{-1}$] & [km~s$^{-1}$]   &        \\
\hline
\multicolumn{8}{l}{\it IRAM~30-m:}\\
\hline
02/16.49--16.59 & 0.136 & 0.978 &  50  & HCN(3-2) & $4.146\pm0.061$ & $-0.02\pm0.03$ & 1.8\arcsec \\
                & 0.136 & 0.978 &  19  & HCN(3-2) & $2.173\pm0.112$ & $-0.01\pm0.11$ & 5.1\arcsec \\
02/17.48--17.52 & 0.108 & 0.985 &  19  & HCN(3-2) & $3.558\pm0.069$ & $+0.01\pm0.04$ & 1.8\arcsec \\
02/17.48--17.50 & 0.109 & 0.985 &   9  & HCN(3-2) & $1.819\pm0.140$ & $-0.18\pm0.17$ & 5.7\arcsec \\

02/17.55--17.66 & 0.106 & 0.987 &  85  & HCN(1-0) & $0.116\pm0.021$ & $+0.06\pm0.32$ & 1.8\arcsec \\

02/16.49--16.59 & 0.135 & 0.978 &  29  & HNC(3-2) & $0.648\pm0.113$ & $+0.09\pm0.35$ & 5.5\arcsec \\
02/16.49--16.66 & 0.134 & 0.978 &  85  & HNC(3-2) & $0.914\pm0.060$ & $+0.25\pm0.13$ & 2.8\arcsec \\

02/16.49--16.59 & 0.135 & 0.978 &  27  & CS(3-2) & $0.242\pm0.054$ & $+0.57\pm0.42$ & 5.5\arcsec \\
02/16.49--16.66 & 0.134 & 0.978 &  87  & CS(3-2) & $0.336\pm0.026$ & $+0.05\pm0.13$ & 2.8\arcsec \\
02/17.48--17.52 & 0.108 & 0.985 &  22  & CS(3-2) & $0.373\pm0.046$ & $-0.34\pm0.21$ & 1.7\arcsec \\
02/17.48--17.50 & 0.109 & 0.985 &   6  & CS(3-2) & $0.180\pm0.085$ &                & 5.6\arcsec \\

02/16.49--17.50 & 0.130 & 0.980 &  22  & CH$_3$CN(8-7)$^1$ & $<0.126$ & & 2.5\arcsec \\

02/17.61--17.66 & 0.106 & 0.987 &  35  & CO(2-1) & $<0.161$ & & 2.8\arcsec \\

02/16.54--17.66 & 0.120 & 0.982 &  80  & OCS(12-11) & $<0.029$ & & 2.1\arcsec \\
        &   &   &  & HC$_3$N(16-15) & $<0.031$ & & 2.1\arcsec \\
        &   &   &  & H$_2$CO($2_{02}-1_{01}$) & $0.032\pm0.012$ & & 2.1\arcsec \\

02/17.55--17.60 & 0.107 & 0.986 & 50 & H$_2$CO($3_{12}-2_{11}$) & $<0.136$ & & 1.0\arcsec \\

02/16.60--17.60 & 0.120 & 0.982 & 107 & CH$_3$OH(3,3-3,2)A$^{\pm~2}$ & $0.072\pm0.045$ & & 1.8\arcsec \\
        &   &   &  & CH$_3$OH(4,3-4,2)A$^\pm$      & $0.213\pm0.046$ & $+0.20\pm0.35$ & \\
        &   &   &  & CH$_3$OH(5,3-5,2)A$^\pm$      & $0.092\pm0.044$ & & \\
        &   &   &  & CH$_3$OH(6,3-6,2)A$^\pm$      & $0.182\pm0.047$ & $+0.31\pm0.48$ & \\
        &   &   &  & CH$_3$OH(7,3-7,2)A$^\pm$      & $0.198\pm0.041$ & $-0.28\pm0.29$ & \\
        &   &   &  & CH$_3$OH(8,3-8,2)A$^\pm$      & $0.057\pm0.041$ & & \\

02/16.60--17.60 & 0.120 & 0.982 & 107 & SO($5_6-4_5$) & $<0.071$ & & 1.8\arcsec \\

02/17.61--17.66 & 0.106 & 0.987 &  35  & SiO(6-5) & $<0.190$ & & 1.7\arcsec \\

\hline
\end{tabular}
\end{center}
$^1$: sum of CH$_3$CN(8,0-7,0), CH$_3$CN(8,1-7,1), CH$_3$CN(8,2-7,2) and CH$_3$CN(8,3-7,3) lines.; \\
$^2$: sums of twin lines;
\end{table*}
}

\onltab{4}{
\begin{table*}
\caption[]{Molecular observations in comet C/2006~P1 (Mc Naught)}\label{tabobs06p1}
\begin{center}
\begin{tabular}{cccrlccr}
\hline
UT date (2007) & $<r_{h}>$ & $<\Delta>$ & Int. time & Line & $\int T_bdv$    & Velocity offset & offset \\[0cm]
[mm/dd.dd--dd.dd] & [AU] & [AU]  &    [min]  &      & [K~km~s$^{-1}$] & [km~s$^{-1}$]   &        \\
\hline
\multicolumn{8}{l}{\it IRAM~30-m:}\\
\hline
01/15.65        & 0.207 & 0.817 &  0.1 & HCN(3-2) & $83.74\pm 7.63$ & $-0.27\pm0.16$ & 2.6\arcsec \\
        & 0.207 & 0.817 &  0.2 & HCN(3-2) & $57.66\pm 5.82$ & $+0.33\pm0.18$ & 5.0\arcsec \\
        & 0.207 & 0.817 &  2.0 & HCN(3-2) & $31.78\pm 3.07$ & $+0.40\pm0.18$ & 7.4\arcsec \\
        & 0.207 & 0.817 &  0.3 & HCN(3-2) & $22.89\pm 5.08$ & $-0.21\pm0.38$ & 8.9\arcsec \\
        & 0.207 & 0.817 &  0.8 & HCN(3-2) & $13.76\pm 3.32$ & $+0.33\pm0.44$ &12.1\arcsec \\
01/16.53        & 0.228 & 0.821 &  8.0 & HCN(3-2) & $48.45\pm 0.42$ & $-0.07\pm0.02$ & 4.9\arcsec \\
01/16.54        & 0.228 & 0.821 &  8.0 & HCN(3-2) & $63.65\pm 0.39$ & $-0.09\pm0.01$ & 2.1\arcsec \\
01/16.55        & 0.228 & 0.821 & 20.0 & HCN(3-2) & $37.47\pm 0.26$ & $-0.06\pm0.02$ & 6.5\arcsec \\
01/16.54        & 0.228 & 0.821 &  4.0 & HCN(3-2) & $27.28\pm 0.57$ & $-0.03\pm0.05$ & 8.2\arcsec \\
01/16.58        & 0.229 & 0.822 & 12.0 & HCN(3-2) & $47.89\pm 0.28$ & $-0.03\pm0.01$ & 4.7\arcsec \\
01/16.59        & 0.229 & 0.822 &  8.0 & HCN(3-2) & $16.60\pm 0.33$ & $+0.04\pm0.05$ & 9.3\arcsec \\
01/17.55        & 0.255 & 0.832 &  2.0 & HCN(3-2) & $35.91\pm 0.59$ & $-0.05\pm0.04$ & 3.3\arcsec \\
01/17.54        & 0.255 & 0.832 & 10.0 & HCN(3-2) & $21.81\pm 0.28$ & $-0.02\pm0.03$ & 5.3\arcsec \\
01/17.56        & 0.255 & 0.833 & 31.0 & HCN(3-2) & $15.58\pm 0.16$ & $+0.04\pm0.02$ & 7.1\arcsec \\
01/17.55        & 0.255 & 0.832 &  5.0 & HCN(3-2) & $14.44\pm 0.40$ & $-0.02\pm0.06$ & 8.5\arcsec \\
01/17.58        & 0.256 & 0.833 & 20.0 & HCN(3-2) & $ 6.72\pm 0.22$ & $-0.04\pm0.07$ &10.7\arcsec \\
01/17.58        & 0.256 & 0.833 & 12.0 & HCN(3-2) & $ 5.69\pm 0.27$ & $-0.06\pm0.11$ &12.4\arcsec \\

01/16.54        & 0.228 & 0.821 & 20.0 & HCN(1-0) & $1.723\pm0.057$ & $+0.16\pm0.05$ & 3.9\arcsec \\
01/16.55        & 0.228 & 0.821 & 16.0 & HCN(1-0) & $1.505\pm0.066$ & $+0.02\pm0.06$ & 6.8\arcsec \\
01/16.54        & 0.228 & 0.821 &  4.0 & HCN(1-0) & $1.574\pm0.131$ & $-0.13\pm0.11$ & 8.1\arcsec \\
01/16.59        & 0.229 & 0.822 & 12.0 & HCN(1-0) & $1.920\pm0.031$ & $-0.01\pm0.02$ & 4.8\arcsec \\
01/16.59        & 0.229 & 0.822 &  8.0 & HCN(1-0) & $1.261\pm0.037$ & $-0.08\pm0.03$ &10.2\arcsec \\

01/17.55        & 0.255 & 0.832 &  2.0 & HNC(3-2) &  $4.05\pm 0.55$ & $-0.06\pm0.18$ & 3.7\arcsec \\
01/17.54        & 0.255 & 0.832 &  9.0 & HNC(3-2) &  $3.16\pm 0.24$ & $+0.03\pm0.09$ & 4.7\arcsec \\
01/17.56        & 0.255 & 0.833 & 23.0 & HNC(3-2) &  $1.53\pm 0.16$ & $+0.11\pm0.12$ & 6.7\arcsec \\
01/17.54        & 0.255 & 0.832 & 14.0 & HNC(3-2) &  $1.52\pm 0.20$ & $+0.13\pm0.16$ & 9.1\arcsec \\
01/17.54        & 0.255 & 0.832 &  8.0 & HNC(3-2) &  $0.52\pm 0.26$ &                &11.5\arcsec \\

01/17.55        & 0.255 & 0.832 &  2.0 & CS(3-2)  &  $2.84\pm 0.20$ & $+0.07\pm0.08$ & 3.4\arcsec \\
01/17.54        & 0.255 & 0.832 &  8.0 & CS(3-2)  &  $2.47\pm 0.10$ & $-0.04\pm0.04$ & 4.7\arcsec \\
01/17.56        & 0.255 & 0.833 & 24.0 & CS(3-2)  &  $1.53\pm 0.06$ & $-0.03\pm0.04$ & 6.8\arcsec \\
01/17.54        & 0.255 & 0.832 & 13.0 & CS(3-2)  &  $1.17\pm 0.08$ & $-0.09\pm0.08$ & 9.1\arcsec \\
01/17.54        & 0.255 & 0.832 &  9.0 & CS(3-2)  &  $1.01\pm 0.09$ & $+0.11\pm0.10$ &11.4\arcsec \\

01/17.51--17.58 & 0.255 & 0.832 & 47 & CH$_3$CN(8,0-7,0) & $0.148\pm0.040$ & $+0.45\pm0.20$ & 6.9\arcsec \\
                &       &       &    & CH$_3$CN(8,1-7,1) & $0.080\pm0.044$ & $-0.60\pm0.47$ & \\
                &       &       &    & CH$_3$CN(8,2-7,2) & $0.004\pm0.039$ &    &  \\
                &       &       &    & CH$_3$CN(8,3-7,3) & $0.104\pm0.042$ & $-1.01\pm0.62$   &  \\
                &       &       &    & Sum of the 4 lines & $0.334\pm0.083$ &   &  \\
01/16.52--16.60 & 0.228 & 0.821 & 48 & CH$_3$OH(1,0-1,-1)E & $0.203\pm0.050$ &       & 4.9\arcsec\\
                &       &       &    & CH$_3$OH(2,0-2,-1)E & $0.160\pm0.047$ & $-0.52\pm0.44$ & \\
                &       &       &    & CH$_3$OH(3,0-3,-1)E & $0.309\pm0.050$ &       & \\
                &       &       &    & CH$_3$OH(4,0-4,-1)E & $0.327\pm0.048$ & $-0.06\pm0.21$ & \\
                &       &       &    & CH$_3$OH(5,0-5,-1)E & $0.305\pm0.047$ & $+0.14\pm0.22$ & \\
                &       &       &    & CH$_3$OH(6,0-6,-1)E & $0.347\pm0.047$ & $-0.05\pm0.20$ & \\
                &       &       &    & CH$_3$OH(7,0-7,-1)E & $0.297\pm0.042$ & $-0.24\pm0.20$ & \\
01/16.54--16.60 & 0.228 & 0.821 &  8 & CH$_3$OH($J,0-J,-1$)E$^1$ & $0.857\pm0.198$ & $-0.60\pm0.35$ & 8.9\arcsec\\
01/17.51--17.58 & 0.255 & 0.832 & 48 & CH$_3$OH(1,0-1,-1)E & $0.024\pm0.035$ &       & 6.7\arcsec\\
                &       &       &    & CH$_3$OH(2,0-2,-1)E & $-0.021\pm0.039$ &      & \\
                &       &       &    & CH$_3$OH(3,0-3,-1)E & $0.081\pm0.035$ &       & \\
                &       &       &    & CH$_3$OH(4,0-4,-1)E & $0.028\pm0.042$ & $+0.64\pm0.60$ & \\
                &       &       &    & CH$_3$OH(5,0-5,-1)E & $0.095\pm0.040$ & $-0.47\pm0.46$ & \\
                &       &       &    & CH$_3$OH(6,0-6,-1)E & $0.103\pm0.045$ & $-0.17\pm0.47$ & \\
                &       &       &    & CH$_3$OH(7,0-7,-1)E & $0.082\pm0.040$ & $-0.55\pm0.40$ & \\
                &       &     && CH$_3$OH($J,0-J,-1$)E$^1$ & $0.380\pm0.079$ & $-0.40\pm0.29$ & 6.7\arcsec\\

01/16.52--16.60 & 0.228 & 0.821 & 60 & H$_2$CO($3_{12}-2_{11}$) & $0.334\pm0.050$ & $-0.18\pm0.20$ & 6.1\arcsec\\
01/16.52--16.59 & 0.228 & 0.821 & 36 & HDO($3_{12}-2_{11}$) & $0.227\pm0.064$ & $-0.68\pm0.40$ & 4.2\arcsec \\
01/15.60        & 0.207 & 0.817 & 5.0 & CO(2-1)  &  $<2.31$ &               &12.1\arcsec \\
01/17.59--17.61 & 0.256 & 0.833 & 24  & CO(2-1)  &  $0.30\pm 0.09$ & $-0.07\pm0.43$ &11.2\arcsec \\
01/17.59--17.61 & 0.256 & 0.833 & 24  & HCO+(1-0)&  $0.36\pm 0.05$ & $-0.05\pm0.27$ &10.9\arcsec \\
\hline
\end{tabular}
\end{center}
$^1$: sum of CH$_3$OH lines $J,K$ = (4,0-4,-1)E, (5,0-5,-1)E, (6,0-6,-1)E and (7,0-7,-1)E.
\end{table*}
}

The estimated visual magnitudes of these comets at perihelion 
were $m_1\approx+3$, $m_1\approx-1.5$, and $m_1\approx-5$ for 
C/2002~X5, C/2002~V1, and C/2006~P1, respectively.
 Correcting for the $-2$ magnitude surge in brightness of C/2006~P1
caused by forward scattering \citep{Mar07}, and using the correlation law
between heliocentric magnitude and $Q_{\rm H_2O}$ of \citet{Jor08},
this would imply water production rates of 1, 11, and 26$\times10^{30}$
\mols respectively. The comparison to measured production rates
in Table~\ref{tabqh2o} suggests that C/2002~V1 was a more dusty comet
than the two others since, unlike C/2002~X5 and C/2006~P1, the actual 
outgassing rate being much lower than
the value ($11\times10^{30}$~\mols) inferred from visual magnitudes.

\begin{figure*}[ht]
\centering
\resizebox{\hsize}{!}{\includegraphics[angle=270]{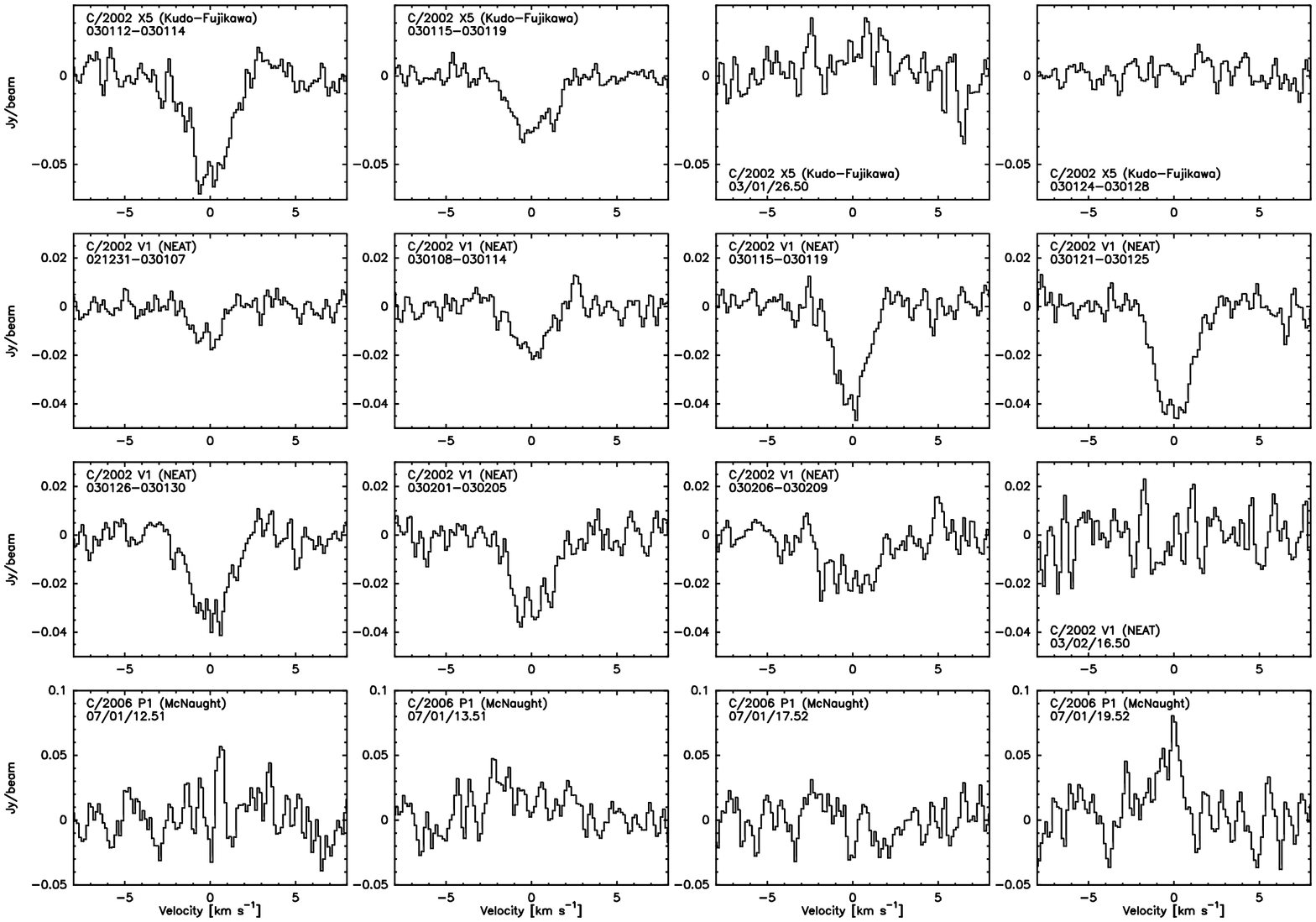}}
\caption{Nan\c{c}ay spectra of OH 18-cm lines (average of 1665 and 
1667 MHz transitions) corresponding to observations listed in 
Table~\ref{tabqh2o}.}
\end{figure*}\label{fignancay}

\begin{table*}
\caption[]{Water production rates}\label{tabqh2o}
\begin{center}
\begin{tabular}{lllcrc}
\hline
UT date    & $<r_{h}>$  &  Observatory    & Line intensity & $Q_{\rm H_2O}$ & Ref.\\[0cm]
[mm/dd.dd] &   [AU]     &  and line       &  [mJy~\kms]      & [\mols] & \\
\hline
\multicolumn{6}{l}{C/2002~X5 (Kudo-Fujikawa): (2003)}\\
\hline
01/12.5--14.5 & 0.55  & Nan\c{c}ay OH 18-cm     & $-156\pm7$   & $1.5\pm0.1\times10^{29}$ &  \\
01/15.5--19.5 & 0.45  & Nan\c{c}ay OH 18-cm     & $-91\pm5$    & $1.8\pm0.1\times10^{29}$ &  \\
01/26.50      & 0.214 & Nan\c{c}ay OH 18-cm     & $47\pm15$    & $85\pm27\times10^{29}$ &   \\
01/25.5--30.5 & 0.205 & Nan\c{c}ay OH 18-cm     & $\approx32\pm 8$     & $\approx64\pm16\times10^{29}$ &   \\
01--02        & $< 1$ & SOHO-SWAN           &  -    & $0.56\times10^{29}\times r_h^{-2.0}$ & [3]\\
01/27.92      & 0.195 & SOHO-UVCS Lyman~$\alpha$  & -       & $5.5\times10^{29}$ & [1] \\
01/28.13      & 0.194 & SOHO-UVCS Lyman~$\alpha$  & -       & $9.0\times10^{29}$ & [1] \\
01/28.79      & 0.190 & SOHO-UVCS Lyman~$\alpha$  & -       & $13.0\times10^{29}$ & [1] \\
01/29.17      & 0.190 & SOHO-UVCS Lyman~$\alpha$  & -       & $27.0\times10^{29}$ & [1] \\
03/12.5       & 1.178 & Odin H$_2$O 557GHz      & $879\pm79^1$    & $0.21\pm0.01\times10^{29}$  & [2] \\
\hline
\multicolumn{6}{l}{C/2002~V1 (NEAT): (2003)}\\
\hline
01/00--07     & 1.30  & Nan\c{c}ay OH 18-cm     &  $-25\pm4$   & $0.15\pm0.02\times10^{29}$ &  \\
01/08--14     & 1.15  & Nan\c{c}ay OH 18-cm     &  $-45\pm4$   & $0.24\pm0.02\times10^{29}$ &  \\
01/15--19     & 1.01  & Nan\c{c}ay OH 18-cm     &  $-81\pm5$   & $0.46\pm0.03\times10^{29}$ &  \\
01/21--25     & 0.87  & Nan\c{c}ay OH 18-cm     & $-110\pm6$   & $0.55\pm0.03\times10^{29}$ &  \\
01/26--30     & 0.73  & Nan\c{c}ay OH 18-cm     &  $-93\pm5$   & $0.59\pm0.03\times10^{29}$ &  \\
02/01--05     & 0.59  & Nan\c{c}ay OH 18-cm     &  $-85\pm6$   & $0.70\pm0.06\times10^{29}$ &  \\
02/06--09     & 0.45  & Nan\c{c}ay OH 18-cm     &  $-66\pm7$   & $1.7\pm0.2\times10^{29}$ &  \\
02/16.50      & 0.136 & Nan\c{c}ay OH 18-cm     &  $-2\pm13$   & $<150\times10^{29}$ &  \\
01--03        & $< 1$ & SOHO-SWAN    &  -    & $0.94\times10^{29}\times r_h^{-2.1}$ & [3] \\
\hline
\multicolumn{6}{l}{C/2006~P1 (McNaught): (2007)}\\
\hline
01/12.51      & 0.171 & Nan\c{c}ay OH 18-cm     & $45\pm23$     & $<270\times10^{29}$ & \\
01/13.51      & 0.173 & Nan\c{c}ay OH 18-cm     & $111\pm15$    & $400\pm60\times10^{29}$ & \\
01/17.52      & 0.254 & Nan\c{c}ay OH 18-cm     & $-9\pm23$     & $<100\times10^{29}$ & \\
01/19.52      & 0.311 & Nan\c{c}ay OH 18-cm     & $128\pm26$    & $80\pm20\times10^{29}$ & \\
\hline
\end{tabular}
\end{center}
$^1$: line integrated intensity in mK~\kms; \\[0cm]
[1]: \citet{Pov03};
[2]: \citet{Biv07a};
[3]: \citet{Com08}
\end{table*}


\section{Analysis of IRAM data}
    Molecular production rates were derived using
models of molecular excitation and radiation transfer
\citep{Biv99,Biv00,Biv06a}. The excitation model
of CH$_3$OH was updated. The computation of the partition 
function considers now the first torsional state, 
which is populated significantly in the hot atmospheres of
the comets studied in this paper. For these productive comets
observed at small $r_h$, a significant fraction of the observed
molecules are photodissociated before leaving the
collision-dominated region. Thus, the determination of the gas
kinetic temperature (which controls the rotational level
populations in the collision zone) and strong constraints on
molecular lifetimes were essential. As shown below, line shapes and
brightness distributions obtained from coarse maps provide
information on the gas outflow velocity and molecular lifetimes.

\subsection{Gas temperature}
Table~\ref{tabtemp} summarizes rotational temperatures deduced
from relative line intensities. Using our excitation models, we
then constrained the gas temperature (Table~\ref{tabtemp})
following the methods outlined in, e.g., \citet{Biv99}. Several data
are indicative of relatively high temperatures, which are
difficult to measure because the rotational population
is spread over many levels making individual lines weaker. Hence,
the uncertainties in derived values are high. We note that the values
derived for C/2002~X5 from CH$_3$OH lines do not reflect the
marginal detection of some lines. The inconsistency between the
different measurements, especially from CH$_3$OH and HCN lines in
comet C/2006~P1, is possibly related to differences in collision
cross-sections and temperature variations in the coma.

Gas temperature laws as a function $r_h$ were obtained for comets
C/1995~O1 (Hale-Bopp) \citep{Biv02b}, C/1996~B2 (Hyakutake) 
\citep{Biv99}, and 153P/Ikeya-Zhang \citep{Biv06a}. 
The laws ($100r_h^{-1.1}$ for Hale-Bopp and $60r_h^{-0.9}$ 
for other comets on average), extrapolate
to $T=200$--1000~K for  $r_h = 0.25$--0.11 AU. Photolytic
heating increases with decreasing $r_h$ and the increasing 
production rate of water. Hence, we
expect higher temperatures for the most productive comet C/2006~P1.
\citet{Com88} predicted a maximum temperature on the order of 
500--900~K  for comet Kohoutek at 0.25--0.14 AU, but temperatures 
are generally below 300~K at distances from the nucleus
where molecules are not photodissociated. Given also that the
HCN $J$(3--2) $v_2=1$ line at 265852.709 MHz is not detected,
we estimate that temperatures are lower than 300~K
below $r_h$=0.25 AU.

For C/2002~X5, we adopted the gas kinetic temperature values of
50~K, 180~K, and 40~K for mid-January, end of January, and mid-March
all in 2003, respectively. For C/2002~V1 and C/2006~P1, we used 150~K
and 300~K, respectively. We later discuss (in Sect. 5.1) the influence 
of the assumed temperature on the inferred production rates.

\begin{table*}
\caption[]{Gas temperature measurements}\label{tabtemp}
\begin{center}
\begin{tabular}{ccclcc}
\hline
UT date    & $<r_{h}>$ &   offset  &  Lines & Rotational temperature & Gas temperature\\[0cm]
[mm/dd.dd] &   [AU]    & [\arcsec] &            &   [K]     & [K]\\
\hline
\multicolumn{6}{l}{C/2002~X5 (Kudo-Fujikawa): (2003)}\\
\hline
01/13.61 & 0.553 & 3.5 & CH$_3$OH 157~GHz       & $38\pm10$         & $30\pm10$ \\
01/26.54 & 0.213 & 3.0 & CH$_3$OH 157~GHz       & $320^{+\infty}_{-170}$    & $175^{+\infty}_{-90}$ \\
01/26.54 & 0.213 & 1.6 & CH$_3$CN 147~GHz       & $1500^{+\infty}_{-1305}$ & $>190$ \\
01/26.54 & 0.213 & 1.0 & HCN$J$(3--2),$v_2=1/v_2=0$ & $<245$        & $<530^1$ \\
03/12.72 & 1.184 & 3.0 & CH$_3$OH 157~GHz       & $85^{+68}_{-25}$  & $70^{+50}_{-20}$ \\
\hline
\multicolumn{6}{l}{C/2002~V1 (NEAT): (2003)}\\
\hline
02/17.0  & 0.120 & 1.9 & HCN$J$(3--2),$v_2=1/v_2=0$     & $<285$        & $<574^1$ \\
02/17.25 & 0.116 & 1.6 & CH$_3$OH 252~GHz       & $83^{+77}_{-27}$  & $80^{+80}_{-30}$ \\
02/17.3  & 0.110 & 1.6 & CH$_3$OH 252~GHz+$20_3$A$^{+-}$ & $228^{+92}_{-51}$ & $220^{+90}_{-50}$ \\
02/17.6  & 0.108 & 1.8 & HCN$J$(3--2)/(1--0)    & $54^{+60}_{-22}$  & $80^{+95}_{-35}$ \\
\hline
\multicolumn{6}{l}{C/2006~P1 (McNaught): (2007)}\\
\hline
01/16.56 & 0.228 & 3   & HCN$J$(3--2)/(1--0)    & 299$^{+\infty}_{-204}$ $^2$    & $>375$ $^2$ \\
01/16.56 & 0.228 & 5   & HCN$J$(3--2)/(1--0)    & $>168$ $^2$       & $>600$ $^2$ \\
01/16.55 & 0.229 & 5.9 & CH$_3$OH 157~GHz       & $149^{+\infty}_{-77}$     & $110^{+\infty}_{-50}$ \\
01/16.56 & 0.229 & 4.9 & CH$_3$OH 157~GHz       & $88^{+32}_{-19}$  & $70^{+20}_{-10}$ \\
01/17.5  & 0.255 & 6.5 & HCN$J$(3--2),$v_2=1/v_2=0$ & $<278$        & $<677^1$ \\
01/17.55 & 0.255 & 6.7 & CH$_3$OH 157~GHz       & $>130$        & $>90$ \\
01/17.55 & 0.255 & 6.9 & CH$_3$CN 147~GHz       & $78^{+106}_{-28}$     & $80^{+110}_{-30}$ \\
\hline
\end{tabular}
\end{center}
$^1$ A significant fraction of the molecules are outside the collisional 
region and a higher gas temperature is needed to populate the
$v=1$ level inside the collision zone.

$^2$ A 10\% calibration uncertainty in each line is assumed.
\end{table*}

\subsection{Screening of photolysis by water molecules}

Given high water-production rates ($> 10^{30}$--10$^{31}$ \mols),
self-shielding against photodissociation by solar UV radiation
is significant for these comets. As a consequence,
molecules can have longer lifetimes on the night side, and reduced
H$_2$O photolysis may limit gas acceleration. We studied the
screening of photolysis using simplified assumptions: isotropic
outgassing at a constant expansion velocity and an infinite lifetime
for the screening molecules (i.e., we assumed that their scale
length is larger than the size of the optically thick region).
According to \citet{Nee85}, when considering \citet{Lee86} and 
\citet{Lee84}, the OH photoabsorption cross-section does not differ
much from the H$_2$O one: it peaks close to the Lyman~$\alpha$ 
wavelength and is slightly stronger (1--2 times). On the other hand, the
cometary hydrogen may not absorb significantly the Solar Lyman~$\alpha$ spectral line. 
The comet Lyman~$\alpha$ line is too narrow (velocity dispersion of 20~\kms) 
to absorb significantly the solar emission ($\approx$180~\kms~ line width).
Consequently, if the sum of the scalelengths of H$_2$O and OH is 
larger than the distances from the nucleus considered hereafter, the
infinite lifetime assumption should not underestimate the screening effect.

\begin{figure*}
 \sidecaption
 \includegraphics[angle=270,width=12cm]{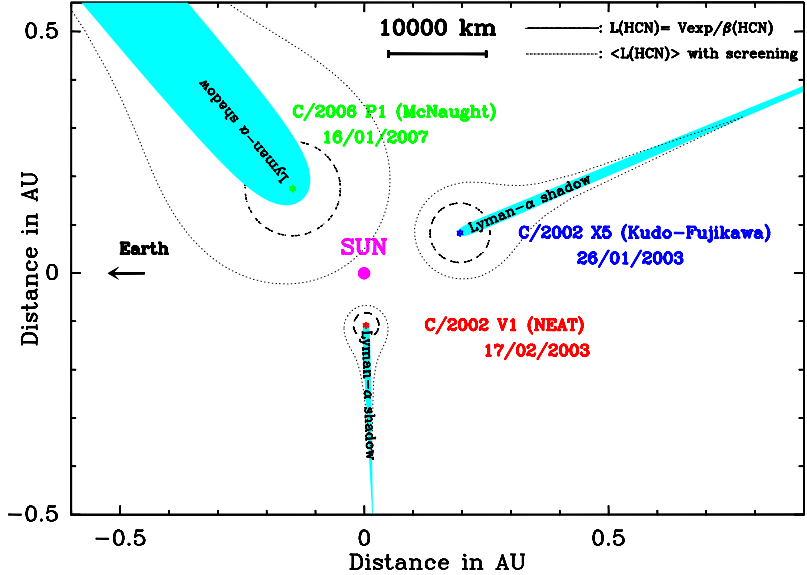}
 \caption{Sketch of the Earth-Comet-Sun plane, with the positions of 
each of the three comets at the epoch of the respective observations.
 In addition (at a different scale given by the 10000~km bar), 
three regions in the comae are plotted: in the shaded area, the region opaque to 
Lyman~$\alpha$ (i.e. where opacity at 121.6~nm towards the Sun is larger 
than 1.0); in dashed line, the spherical region whose radius corresponds to the 
Haser scale length of HCN (assuming a constant mean value for the 
expansion velocity); within the dotted lines,
the region where less than 1-1/e of the HCN molecules are photodissociated
($\approx$ equivalent to its scale length) when the screening is taken into account.}
\label{figgeom}
\end{figure*}

In cylindrical coordinates, with the vertical z-axis pointing
towards the Sun, a point in the coma has coordinates ($\theta$,
$\rho = r\sin(\phi)$, $z_r = r\cos(\phi)$), where $r$ is the distance to
the nucleus and $\phi$ the co-latitude angle.
The photodissociation rate of a molecule M, characterized
by its photodissociation absorption cross-section 
$\sigma_{\rm M}(\lambda)$, in a radiation field $F(\lambda)$
(in photons~m$^{-2}$~s$^{-1}$~nm$^{-1}$) is given by
 
\begin{equation}
  \beta_M = \int_{\lambda}\sigma_{\rm M}(\lambda)F(\lambda)d\lambda.
  \label{eqbeta}
\end{equation}
The problem is symmetric around the z-axis. At ($\rho$, $z_r$),
the solar flux at wavelength $\lambda$ will be attenuated to
\begin{equation}
  F(\lambda,\rho,\phi) = F_0(\lambda)\exp[-\tau(\lambda,\rho,\phi)]
\end{equation}
because of the optical thickness along the comet-sun axis.
We consider water as the major molecule responsible for the opacity,
which is connected to its absorption cross-section  
$\sigma_{\rm H_2O}(\lambda)$ by
\begin{equation}
  \tau(\lambda,\rho,\phi) = \sigma_{\rm H_2O}(\lambda)\int_{z_r}^{\infty} n_{\rm
  H_2O}(\rho,z)dz.
\end{equation}
The photons absorbed are mostly those responsible for photodissociation 
and not fluorescence. 
Photodissociation of molecules takes place for $\lambda < 200$~nm and 
a significant solar UV field ($\lambda > 80$ nm).
Using the assumptions for the density, we can integrate the density
along the z-axis to obtain

\begin{eqnarray*}
 \int_{z_r}^{\infty} n_{\rm H_2O}(\rho,z)dz 
& = & \int^{\infty}_{z_r=\rho/\tan(\phi)} \frac{Q_{\rm H_2O}}{4\pi v_{\rm exp}(\rho^2+z^2)}dz  \\
& = & \frac{Q_{\rm H_2O}}{4\pi v_{\rm exp}r}\frac{\phi}{\sin(\phi)}. \\
\end{eqnarray*}
Hence at any point in the coma, we estimate the opacity
\begin{equation}
\tau(\lambda,r,\phi) = <\tau(\lambda,r)>\frac{4\phi}{\pi^2\sin(\phi)}
        = C(\lambda)\frac{\phi}{r\sin(\phi)},
\label{eqtauphi}
\end{equation}
where the average value of the opacity over $4\pi$
steradians at the distance $r$ from the nucleus is
\begin{equation}
<\tau(\lambda,r)> = C(\lambda)\frac{\pi^2}{4}\frac{1}{r}
\label{eqmeantau}
\end{equation}
with
\begin{equation}
C(\lambda) =  \frac{\sigma_{\rm H_2O}(\lambda)Q_{\rm H_2O}}{4\pi v_{\rm exp}}.
\end{equation}
The surface corresponding to an opacity $\tau(\lambda,r,\phi)=1.0$ 
is defined by
\begin{equation}
\label{eq:rthick}
 r_{thick}(\phi) = C(\lambda)\frac{\phi}{sin(\phi)} =
    \frac{\sigma_{\rm H_2O}(\lambda) Q_{\rm H_2O}}{4\pi v_{\rm exp}}\frac{\phi}{\sin(\phi)}.
\end{equation}
According to Eq.~\ref{eq:rthick}, the size of the optically thick
region tends to infinity in the anti-sunward direction
($\phi=\pi$). It has a finite length determined by the apparent
size of the Sun (2.7\degr~at $r_h=0.2$ AU). This region is plotted
in Fig.~\ref{figgeom} for the three comets, considering only
absorption of Lyman~$\alpha$ photons by water molecules with a
cross-section $\sigma_{\rm H_2O}(Ly~\alpha)$ = $15\times10^{-22}$
m$^2$ \citep{Lew83}.

We next consider HCN photodissociation, but similar
equations can be established for other molecules. The effective
HCN photodissociation rate at a point ($r$,$\phi$) in the coma can
then be derived from Eq.~\ref{eqbeta}
\begin{equation}
\beta_{\rm HCN}(r,\phi) = 
\int_\lambda \sigma_{\rm HCN}(\lambda)F_0(\lambda)\exp\left(-\frac{C(\lambda)\phi}{r\sin(\phi)}\right)d\lambda, \\
\end{equation}
where $\sigma_{\rm HCN}(\lambda)$ is the photodissociation
cross-section for HCN. The integration can be divided over  
several wavelength intervals, corresponding to the different 
absorption bands of HCN and H$_2$O
\begin{eqnarray*}
\beta_{\rm HCN}(r,\phi) & = & \sum_{\lambda_i}\beta_{\rm HCN, \lambda_i}(r,\phi) \\
 & = & \beta_{0,\rm HCN}\times\sum_{\lambda_i}x_i\exp\left(-C(\lambda_i)\frac{\phi}{r\sin(\phi)}\right), \\
\end{eqnarray*}
where $x_i$ corresponds to the fraction of the photodissociation rate due to
radiation around the wavelength $\lambda_i$. 
For HCN, 88\% of the contribution comes from 
solar Lyman~$\alpha$, i.e., $x_{\rm Ly~\alpha}=0.88$ \citep{Boc85}.
To ease computations, we make the approximation
\begin{equation}
\sum_{\lambda_i}x_i\exp\left(-\frac{C(\lambda_i)\phi}{r\sin(\phi)}\right) \approx
\exp\left(-\sum_{\lambda_i}\frac{x_i C(\lambda_i)\phi}{r\sin(\phi)}\right).
\end{equation}
This rough assumption is valid for small opacities ($r>r_{thick}$) and if the water
absorption cross-sections are the same at all wavelengths $\lambda_i$. 
Otherwise, it will slightly overestimate the screening effect in the opaque 
region. We then define an effective water absorption cross-section 
for HCN (likewise for the other molecules)
\begin{eqnarray}
<\sigma_{\rm H_2O\rightarrow HCN}> =\sum_{\lambda_i}x_i\sigma_{\rm H_2O}(\lambda_i)
= 13.8\times10^{-22}~{\rm m}^2,
\end{eqnarray}
where most of the contribution comes from Lyman~$\alpha$
\citep[$\sigma_{\rm H_2O}(Ly~\alpha)=15\times10^{-22}$ m$^2$,][]{Lee86,Lew83}.
This approximation was validated by ourselves for HCN: 
we estimated numerically that using this mean value for the screening
cross-section instead of summing over various wavelength intervals yields 
only a $\approx2$\% excess error in the estimate of the increase in
the number of molecules due to screening.

After solving the differential balance equation for the HCN density at the
distance $r$ from the nucleus in the coma, we find that
\begin{equation}
\label{eq:dens} n_{\rm HCN}(r,\phi) = \frac{Q_{\rm HCN}}{4\pi
v_{\rm exp}r^2}\exp\left(-\frac{r\beta_0}{v_{\rm exp}}G(\frac{r}{C_{\rm HCN}\frac{\phi}{\sin(\phi)}})\right)
\label{eqdens}
\end{equation}
with
\begin{equation}
C_{\rm HCN} = \frac{<\sigma_{\rm H_2O\rightarrow HCN}>Q_{\rm H_2O}}{4\pi v_{\rm exp}}.
\end{equation}
The function $G(x) = \frac{1}{x}\int_0^x \exp(-1/t)dt$ is equal to
the exponential integral $G(x)$= E$_2(1/x)$, which can easily be
computed by numerical integration. We note that $G(\infty)=1.0$, so
that Eq.~\ref{eq:dens} gives the classical Haser formula
for negligible opacities.

We compared the production rates determined using the density from
Eq.~\ref{eqdens} to those obtained with the Haser formula. The
largest effect is for comet C/2006~P1 with a $\approx$ 60\%
decrease of the HCN production rate, and $\approx$ 40\% decrease
for CH$_3$OH, CH$_3$CN, or HDO. We did not develop a full 3-D model to
take into account phase angles different from 0\degr~ or 180\degr.
But the comparison between the case of a phase angle of 180\degr~,
where we can simply use Eq.~\ref{eqdens}, and replacing Eq.~\ref{eqtauphi} 
by the averaged value in Eq.~\ref{eqmeantau}, only yields a 3\% difference.

Table~\ref{tabscreen} provides characteristic scalelengths for the
three comets and for comparison C/1996 B2 (Hyakutake) and C/1995 O1 (Hale-Bopp)
observed in 1996--1997. In all cases, the optically thick region 
(at the Lyman~$\alpha$ wavelength) is within the water coma 
($L_{\rm H_2O} >$ $\langle r_{thick}(Ly~\alpha)\rangle$) and if one takes into
account OH, the assumption of an infinite lifetime for the screening
molecules is valid since the coma encompassing  H$_2$O and OH is definitely 
larger than the optically thick region and the HCN coma. One can also note that
for comets C/1996~B2 and C/1995~O1,  $L_{\rm HCN} \gg$ 
$\langle r_{thick}(Ly~\alpha)\rangle$, so that the screening effect 
is negligible.

\begin{table*}
\caption[]{Typical scale lengths affected by screening}\label{tabscreen}
\begin{center}
\begin{tabular}{lrrrrrrr}
\hline
Comet & $<r_{h}>$ & $Q_{\rm H_2O}$ & $v_{\rm exp}$ & $L_{\rm H_2O}$ & $L_{\rm OH}$ & $<r_{thick}(\rm Ly~\alpha)>$ & $L_{\rm HCN}$ \\[0cm]
         &  [AU] &   [$10^{29}s^{-1}$]    & [\kms] & [km]   & [km] & [km] & [km] \\
\hline
C/1996~B2 (Hyakutake) & 0.25 &   5 & 1.60 &  7900 &  14300 &   90 &  6600 \\
C/1995~O1 (Hale-Bopp) & 0.91 & 100 & 1.10 & 70000 & 128000 & 2700 & 58000 \\
C/2002~X5 (Kudo-Fujikawa)& 0.21 &  35 & 1.25 &  3600 &   7200 &  820 &  2900 \\
C/2002~V1 (NEAT)      & 0.12 &  20 & 2.00 &  2000 &   3800 &  300 &  1600 \\
C/2006~P1 (McNaught)  & 0.23 & 300 & 1.50 &  8100 &  14900 & 5900 &  4900 \\
\hline
\end{tabular}
\end{center}
$L_{\rm H_2O}$, $L_{\rm OH}$, $L_{\rm HCN}$: unscreened scale lengths
of H$_2$O, OH, and HCN, taking into account solar activity, 
expansion velocity, and heliocentric distance.\\
$<r_{thick}(\rm Ly~\alpha)>$: mean radius of the optically thick 
Lyman~$\alpha$ envelope ($\frac{\pi^2}{4}C(\rm Ly~\alpha)$).
\end{table*}

\subsection{Line shapes and gas expansion velocity}

We used the line shapes to estimate the gas expansion velocity
following e.g., \citet{Biv99}.  Table~\ref{tabvexp} lists the
velocities ($VHM$) corresponding to the half maximum intensity on
both negative and positive sides of the lines. For each comet, we
used the most reliably detected lines. To infer the gas velocity from the
measured $VHM$, the processes contributing to line broadening
should be considered. For the HCN(3--2) line, we took into account
its hyperfine structure (broadening of about 0.03 \kms~and 0.06
\kms, on the negative and positive sides, respectively). Thermal
broadening affects the widths of the lines, and depends on the
actual gas temperature. When gas expansion velocities are in the range
1--2 \kms, and temperatures are 100--300 K, thermal
dispersion essentially smooths the line shape and widens it by
less than 0.05 \kms. Optically thick lines, such as HCN(3--2) in
C/2006~P1, are broadened by an additional 0.05 \kms~when the
expansion velocity is constant in the coma.

\begin{figure}
\centering
\resizebox{\hsize}{!}{\includegraphics[angle=270]{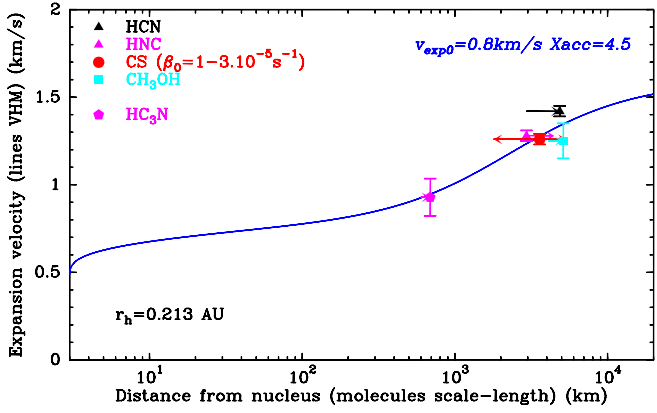}}
\caption{The evolution of the expansion
velocity with distance to the nucleus in the coma of comet C/2002~X5
(Kudo-Fujikawa) on 26.5 January 2003. The measured lines widths have been
plotted with their errorbars at a distance corresponding to the
molecular lifetime. Arrows represent the effect on scalelength of the 
$\beta_0=1$--$3\times10^{-5}$~s$^{-1}$ (Table~\ref{tabvexp}) domain for CS.
For HCN, HNC, and CH$_3$OH, the arrow points to the increase in 
scalelength due to the screening.}
\label{vexp02x5}
\end{figure}

\begin{figure}
\centering
\resizebox{\hsize}{!}{\includegraphics[angle=270]{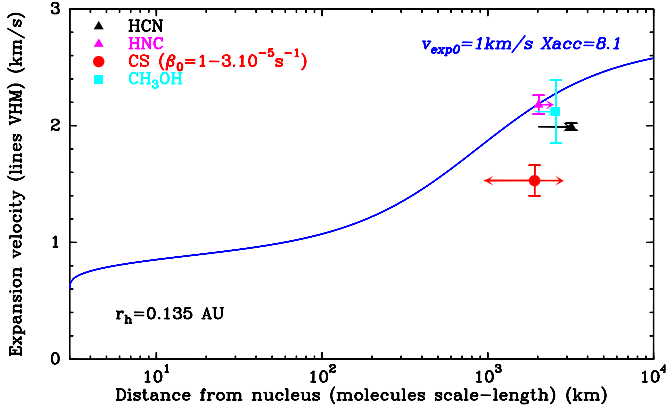}}
\caption{The evolution of the expansion
velocity with distance to the nucleus in the coma of comet C/2002~V1
(NEAT) on 16.5 February 2003. Measurements of line widths from Table~\ref{tabvexp}
are plotted as in Fig.~\ref{vexp02x5}.}
\label{vexp02v1}
\end{figure}

\begin{figure}
\centering
\resizebox{\hsize}{!}{\includegraphics[angle=270]{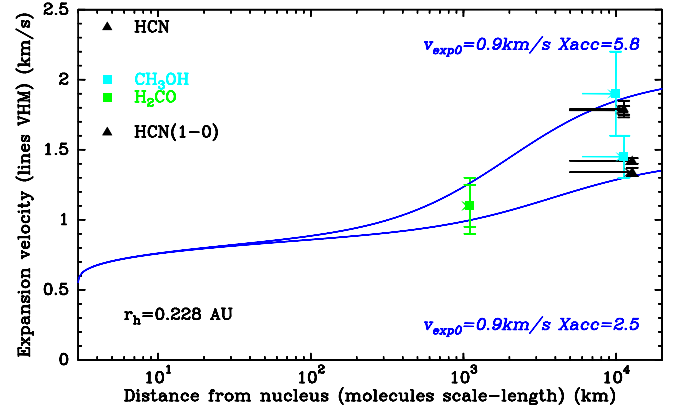}}
\caption{The evolution of the expansion
velocity with distance to the nucleus in the coma of comet C/2006~P1
(McNaught) on 16.5 January 2007. Measurements of line widths from 
Table~\ref{tabvexp} are plotted as in Fig.~\ref{vexp02x5}. 
The upper values ($v_{\rm exp}(r)$ and $VHM$) correspond to the $\sim$dayside
larger velocities ($v>0$).}
\label{vexp06p1-16}
\end{figure}

\begin{figure}
\centering
\resizebox{\hsize}{!}{\includegraphics[angle=270]{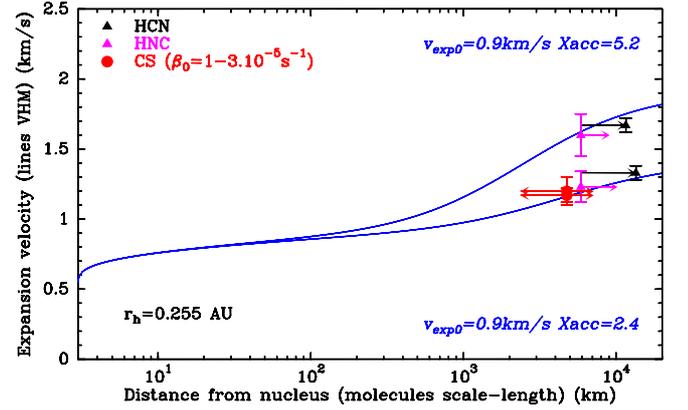}}
\caption{The evolution of the expansion
velocity with distance to the nucleus in the coma of comet C/2006~P1
(McNaught) on 17.5 January 2007. Scales as in Fig.~\ref{vexp06p1-16} with 
upper values for dayside data.}
\label{vexp06p1-17}
\end{figure}

At heliocentric distances $r_h<0.25$ AU, the photodissociation
lifetimes of HCN, HNC, CH$_3$OH, CS, HC$_3$N, and H$_2$CO are shorter
than 1~h. In most cases, the beam size (5000 to 14000~km at the
comet for C/2002~X5, C/2002~V1, and C/2006~P1 near perihelion) is
larger than the molecular scale-length. Therefore, velocity
measurements pertain to nucleus distances fixed by the molecular
lifetimes and the line widths are mostly representative of the
expansion velocity in the coma at a distance equal to the molecular
scale-length. Figures~\ref{spec02x5-5lines}--\ref{spec06p1-4lines}
show that the molecular lines have different widths. The measured
``$VHM$s'' velocities are plotted versus the molecular 
scale-lengths in Figs.~\ref{vexp02x5}--\ref{vexp06p1-17}. In the
following sections, we constrain the evolution of the gas
expansion velocity with distance to nucleus by fitting the
measured $VHMs$, taking into account all sources of line
broadening.

\subsubsection{Outgassing pattern and anisotropy}

    Lines of C/2002~X5 detected around perihelion, and those of
C/2006~P1, show an asymmetric shape, while C/2002~V1 lines are
relatively symmetric.

For C/2002~X5, we likely observed a jet or outburst in progress on
the night side. All the lines are strongly red-shifted (+0.25
\kms~on average, Table~\ref{tabobs02x5}) suggesting gas outflow in
a preferential direction, independent of acceleration effects.
The phase angle was small (26\deg, Table~\ref{tabefe}), so the
observed asymmetry  corresponds to preferential outgassing from
the night side. Anisotropy in screening of photodissociation
(Sect~4.2) also contributes to the asymmetry of the lines because
lifetimes are longer on the night side. Considering the lines the
less affected by screening (HC$_3$N, CH$_3$OH), we found that the
outgassing rate is $1.35\pm0.1$ times higher on the night side.
The mean gas velocity is smaller (0.9 versus 1.2 \kms) on the 
dayside of the nucleus. This is most likely owing to shorter
molecular lifetimes on the dayside, the measurements sampling
thereby molecules closer to the nucleus (Sect.4.2).

For C/2006~P1, the phase angle was large when the comet was
observed (120--140\degr, Fig.~\ref{figgeom}). The line velocity
shifts are slightly negative ($-0.05$ \kms,
Table~\ref{tabobs06p1}), which could indicate some excess emission
from the night side. However, evidence of asymmetric gas
acceleration is present. The mean gas velocities inferred from
long-lived molecules (HCN, CH$_3$OH) are 1.7 and 1.3 \kms~on the
rear ($\approx$ Sun facing) and front side of the nucleus,
respectively, but $\approx$ 1.2 \kms~on both sides for species with
shorter lifetimes (CS and H$_2$CO) (Table~\ref{tabvexp}). The less
significant gas acceleration on the night side is likely due to a
reduced photolytic heating caused by the screening of water
photolysis. The small line blue-shifts suggest that the outgassing
rate is barely larger (by a factor of $1.1\pm0.1$) in the day
side, after taking into account screening effects (Sect.~4.2).

In the case of C/2002~V1, the phase angle was close to 90\degr~
(Table~\ref{tabefe}), so that the line shapes are unaffected by
day/night asymmetries.

\subsubsection{Expansion velocity acceleration}
    Since there is evidence of gas acceleration in the coma, we
used a variable expansion velocity in the coma $v_{exp}(r)$, which
is plotted in Figs.~\ref{vexp02x5}--\ref{vexp06p1-17}. This function
reproduces the general shape of gas-dynamic simulations \citep{Com88}.
The parameter $v_{\rm exp}(r)$ 
\footnote{$v_{\rm exp}(r) =  v_{\rm exp0}\times\left[0.6+0.3\sqrt[3]{\log(r/3)}\right]$
+ $v_{\rm exp0}\times\left[0.1~x_{acc}\left(1+\tanh\left(1.3\log(\frac{r(1+x_{acc})}{60000 r_h})
\right) \right) \right]$, with $r$ in km.}
is adjusted to match the observed line widths (Table~\ref{tabvexp})
when computing line shapes. These velocity plots provide the two free parameters 
used to compute $v_{\rm exp}(r)$ and lead to the following comments:

\begin{enumerate}
  \item C/2002~X5 at $r_h=0.21$ AU (Fig.~\ref{vexp02x5}): 
Positive and negative sides $VHM$s were
both fitted using the same function. 
The computed line shapes for HC$_3$N(28--27),
CH$_3$OH($5_0-5_{-1}+6_0-6_{-1}+7_0-7_{-1}$)E, and HCN(3--2),
taking into account screening and a night/day outgassing
rate ratio of 1.35, have exactly the measured width within the errorbars.
Doppler shifts of the lines (+0.095, +0.172, +0.243~\kms, predicted,
respectively) are also in good agreement.

  \item C/2002~V1 (Fig.~\ref{vexp02v1}): 
Asymmetry in the lines (due to day/night outgassing 
asymmetry and screening effect) is neither seen nor expected because of the phase
angle. We have thus averaged the $VHM$ measurements on both sides of the
lines . However, the two days of observations require two 
different velocity profiles, with stronger acceleration on the second 
day closer to the Sun. For this comet, we did not observe short-lived species 
(e.g., HC$_3$N, H$_2$CO, or H$_2$S), so the velocity
profile is poorly constrained close to the nucleus. 
The CS line widths suggest that the CS lifetime is significantly shorter
than the HCN lifetime.

  \item C/2006~P1: Data were divided into four subsets: positive and negative 
velocity sides of the lines on the 16th (Fig.~\ref{vexp06p1-16}) and 17th 
of January 2007 (Fig.~\ref{vexp06p1-17}). The species with the smaller
scale length (H$_2$CO) was not observed on 17 Jan. but we assumed
a similar acceleration in the coma for both days, scaled to the HCN 
lines $VHM$s. A stronger acceleration is present for molecules moving 
in an opposite direction to the observer ($v>0$), which is consistent with 
stronger acceleration towards the dayside (Fig.~\ref{figgeom}: phase angle 
of 120--140\degr $\ll$ 90\degr).
\end{enumerate}

\subsection{Constraints on CS and HNC lifetimes}
\label{sec:cslife}

	We assume that CS is the photodissociation product of CS$_2$.
It forms very close to the nucleus since CS$_2$ photodissociates into
CS in less than 30 seconds at $r_h<0.25$ AU. We assume that there is no 
additional excitation effect to those not modeled in \citet{Biv99} that would 
strongly affect its rotational population mimicking photodissociation. Hence, the
width of CS lines (clearly narrower than HCN lines; Figs~\ref{spec02x5-5lines},
\ref{spec02v1-4lines}, \ref{spec06p1-4lines}, and Table~\ref{tabvexp}) 
can be converted into a scale-length estimate 
using line-shape modeling with $v_{\rm exp}(r)$. 
The scale-length of CS is determined by the gas expansion velocity, 
its photodissociation rate, and the screening effects.
Since the photodissociation process of CS is unknown, we investigated 
$\beta_{0,\rm CS}=1$--$3\times10^{-5}$~s$^{-1}$ for the photodissociation rate 
at 1 AU and $\sigma_{\rm H_2O\rightarrow CS}=$0--12$\times10^{-22}$~m$^2$ 
(i.e. values smaller than or comparable to HCN for the screening cross-section).
Table~\ref{tabbetacs} summarizes the findings. 
There is no unique solution, but grouping measurements comet by comet, the 
closer matches are found for ($\beta_{0,\rm CS}\geq1.7\times10^{-5}$~s$^{-1}$,
$\sigma_{\rm  H_2O\rightarrow CS}\geq6\times10^{-22}$~m$^2$),
($\beta_{0,\rm CS}=2.9\pm0.7\times10^{-5}$~s$^{-1}$, for
$\sigma_{\rm  H_2O\rightarrow CS}=6\times10^{-22}$~m$^2$), and 
($\beta_{0,\rm CS}=4.3\pm1.2\times10^{-5}$~s$^{-1}$,
$\sigma_{\rm  H_2O\rightarrow CS}\geq6\times10^{-22}$~m$^2$) for C/2002~X5, 
C/2002~V1, and C/2006~P1, respectively. The screening of photodissociation is 
necessary to achieve closer agreement between the night- and dayside 
measurements. In addition, the detection of CS at only 0.1 AU from 
the Sun in C/2002~V1 requires a lifetime 
($\beta_{0,\rm CS}\ll 4\times10^{-5}$~s$^{-1}$) long enough
to get realistic abundances (CS/H$_2$O $\ll$ 1\%, this ratio being $<$ 0.2\%
at 1 AU in all comets where it has been measured).
Values $\beta_{0,\rm CS}=2.5\pm0.5\times10^{-5}$~s$^{-1}$ 
and $\sigma_{\rm H_2O\rightarrow CS}=6\times10^{-22}$~m$^2$ 
are consistent with all measurements and are
used to determine the production rates.

For HNC, we assume similar spectroscopic properties
(e.g. UV absorption spectrum) to those of HCN. Slight differences 
in line shapes may underline differences. In C/2002~X5 and C/2006~P1, the 
HNC(3--2) line is slightly narrower than HCN(3--2) suggesting stronger
photodissociation, weaker screening, and/or other destruction
processes in the coma. But this effect is not observed 
in C/2002~V1. A possible interpretation is that HNC is partly
created in the coma by chemical reactions, as suggested by \citet{Rod98}.
This would partly compensate for the shortening of its scale-length.
A parent scale-length of $\approx1300$ km at $r_h=0.2$ AU,  
$\beta_{0,\rm HNC}\approx1.3\times\beta_{0,\rm HCN}$, and
$\sigma_{\rm H_2O\rightarrow HNC}\approx0.9\times\sigma_{\rm H_2O\rightarrow HCN}$ 
would provide the closest agreements between model and data, but the 
constraints are not strong enough to infer a definite conclusion.

\begin{table}
\caption[]{Constraints on the CS photodissociation rate at 1 AU ($\beta_{0\rm CS}$)}\label{tabbetacs}
\begin{center}
\begin{tabular}{lccr}
\hline
$<r_{h}>$  &  $\beta_{0\rm CS}$ & $\sigma_{\rm H_2O\rightarrow CS}$ & lifetime$^*$ \\[0cm]
[AU]     &   [s$^{-1}$]     & [m$^2$]   & at 1AU [s] \\
\hline
\multicolumn{4}{l}{C/2002~X5: $VHM > 0$}\\
\hline
0.213   & $0.65\pm0.15\times10^{-5}$ &  no screening & 154000 \\
0.213   &  $1.1\pm0.3 \times10^{-5}$ &  $3\times10^{22}$ & 108100 \\
0.213   & $1.25\pm0.35\times10^{-5}$ &  $6\times10^{22}$ & 107500\\
0.213   &  $1.4\pm0.3 \times10^{-5}$ &  $9\times10^{22}$ & 106900 \\
0.213   &  $1.6\pm0.3 \times10^{-5}$ & $12\times10^{22}$ & 104300 \\
\hline
\multicolumn{4}{l}{C/2002~X5: $VHM < 0$}\\
\hline
0.213   & $2.7\pm0.7\times10^{-5}$ &   no screening & 37000 \\
0.213   & $2.3\pm0.7\times10^{-5}$ &   $3\times10^{22}$ & 49400 \\
0.213   & $2.2\pm0.7\times10^{-5}$ &   $6\times10^{22}$ & 55300 \\
0.213   & $2.1\pm0.7\times10^{-5}$ &   $9\times10^{22}$ & 61000 \\
0.213   & $2.0\pm0.7\times10^{-5}$ &  $12\times10^{22}$ & 66600 \\
\hline
\multicolumn{4}{l}{C/2002~V1: 16.6 Jan. 2003}\\
\hline
0.136   & $2.8\pm1.0\times10^{-5}$ &  no screening & 36000 \\
0.136   & $3.2\pm1.2\times10^{-5}$ &  $3\times10^{22}$ & 40400 \\
0.136   & $3.8\pm1.6\times10^{-5}$ &  $6\times10^{22}$ & 40800 \\
\hline
\multicolumn{4}{l}{C/2002~V1: 17.5 Jan. 2003}\\
\hline
0.108   & $2.05\pm0.45\times10^{-5}$ &  no screening & 49000 \\
0.108   & $2.45\pm0.6\times10^{-5}$ &  $3\times10^{22}$ & 56400 \\
0.108   & $2.95\pm0.75\times10^{-5}$ &  $6\times10^{22}$ & 57700 \\
\hline
\multicolumn{4}{l}{C/2006~P1: $VHM < 0$}\\
\hline
0.255   & $0.95\pm0.4\times10^{-5}$ &   no screening & 105000 \\
0.255   & $2.05\pm1.1\times10^{-5}$ &   $3\times10^{22}$ &  77000 \\
0.255   &  $3.2\pm2.0\times10^{-5}$ &   $6\times10^{22}$ &  71000 \\
\hline
\multicolumn{4}{l}{C/2006~P1: $VHM > 0$}\\
\hline
0.255   & $4.3\pm2.1\times10^{-5}$ &   no screening & 23000 \\
0.255   & $5.1\pm2.4\times10^{-5}$ &   $3\times10^{22}$ & 36000 \\
0.255   & $6.4\pm3.4\times10^{-5}$ &   $6\times10^{22}$ & 34000 \\
\hline
\end{tabular}
\end{center}
$^*$ Determined from the ratio of the total number of molecules in the coma divided by the 
production rate, scaled as 1/$r_h^2$ to 1 AU.
\end{table}

\begin{table*}
\caption[]{Line widths and photodissociation rates}\label{tabvexp}
\begin{center}
\begin{tabular}{cclccccccr}
\hline
UT date & $<r_{h}>$ & Line & \multicolumn{2}{c}{Beam size} &
                    \multicolumn{2}{c}{$VHM$} & $v_{\rm exp}$ & $\beta_0$ & $L_d$ \\[0cm]
[mm/dd.dd] & [AU] & & [\arcsec] & [km] & [km~s$^{-1}$] & [km~s$^{-1}$]& [km~s$^{-1}$] & [s$^{-1}$] & [km] \\
\hline
\multicolumn{8}{l}{C/2002~X5: (2003)}\\
\hline
01/13.60 & 0.553 & HCN(3--2)$^1$ &  9.6 &  7190 & $-1.09\pm0.07$ & $+1.20\pm0.07$ & 0.90 & $1.7\times10^{-5}$ & 14330 \\
01/26.54 & 0.213 & HCN(3--2)     &  9.4 &  7990 & $-0.97\pm0.03$ & $+1.42\pm0.02$ & 1.25 & $1.92\times10^{-5}$ & 2950--4870$^2$ \\
01/26.54 & 0.213 & HNC(3--2)     &  9.2 &  7820 & $-0.86\pm0.11$ & $+1.28\pm0.03$ & 1.25 & $1.92\times10^{-5}$ & 2950--4500$^2$ \\
01/26.56 & 0.212 & CS(3--2)      & 17.5 & 14880 & $-0.92\pm0.06$ & $+1.26\pm0.03$ & 1.20 & 1--3$\times10^{-5}$& 1800--5400\\
01/26.60 & 0.212 & HC$_3$N(28--27)& 9.7 &  8250 & $-0.73\pm0.18$ & $+0.93\pm0.11$ & 0.90 & $6.8\times10^{-5}$ &   690$^2$ \\
01/26.55 & 0.213 & \multicolumn{8}{l}{CH$_3$OH($(5_0-5_{-1})+(6_0-6_{-1})+(7_0-7_{-1})$E)}  \\
             &   &   & 15.5 & 13170 & $-0.92\pm0.20$ & $+1.25\pm0.10$ & 1.25 & $1.43\times10^{-5}$& 4120-5130$^2$\\
03/12.72 & 1.184 & HCN(3--2)     &  9.6 &  7630 & $-0.65\pm0.04$ & $+0.52\pm0.07$ & 0.65 & $1.6\times10^{-5}$ & 47460 \\
\hline
\multicolumn{8}{l}{C/2002~V1: (2003)}\\
\hline
02/16.55 & 0.136 & HCN(3--2)     &  9.3 &  6600 & $-2.07\pm0.06$ & $+2.26\pm0.04$ & 2.00 & $1.8\times10^{-5}$ &  2080--2950$^2$ \\
02/16.57 & 0.135 & HNC(3--2)     &  9.2 &  6520 & $-2.34\pm0.22$ & $+2.13\pm0.26$ & 2.00 & $1.8\times10^{-5}$ &  2030--2500$^2$ \\
02/16.55 & 0.136 & CS(3--2)      & 17.3 & 12270 & $-1.60\pm0.35$ & $+1.65\pm0.18$ & 1.60 & 1--3$\times10^{-5}$& 1000--3000 \\
02/17.10 & 0.120 & \multicolumn{8}{l}{CH$_3$OH($(7_3-7_2)$A$^{+-}$+$(6_3-6_2)$A$^{+-}$+$(5_3-5_2)$A$^{+-}$+$(4_3-4_2)$A$^{+-}$+$(7_3-7_2)$A$^{-+}$)} \\
     &       &               & 10.0 &  7120 & $-1.96\pm0.34$ & $+2.37\pm0.43$ & 2.00 & $1.38\times10^{-5}$& 2090--2580$^2$ \\
02/17.50 & 0.108 & HCN(3--2)     &  9.3 &  6650 & $-2.07\pm0.07$ & $+2.24\pm0.06$ & 2.00 & $1.8\times10^{-5}$ & 1300--2250$^2$ \\
02/17.50 & 0.108 & CS(3--2)      & 17.3 & 12370 & $-2.60\pm0.60$ & $+1.65\pm0.10$ & 1.60 & 1--3$\times10^{-5}$& 620--1870 \\
02/17.60 & 0.105 & HCN(1--0)     & 26.8 & 19180 &  $-1.3\pm1.1$  &  $+2.9\pm1.1$  & 2.00 & $1.8\times10^{-5}$ & 1230--2200$^2$ \\

\hline
\multicolumn{8}{l}{C/2006~P1: (2007)}\\
\hline
01/16.56 & 0.228 & HCN(3--2)     &  9.3 &  5540 & $-1.42\pm0.02$ & $+1.78\pm0.03$ & 1.50 & $1.60\times10^{-5}$ &  4870--12300$^2$ \\
01/16.56 & 0.228 & HCN(1--0)     & 26.5 & 15780 & $-1.34\pm0.03$ & $+1.79\pm0.06$ & 1.50 & $1.60\times10^{-5}$ &  4870--12300$^2$ \\
01/16.55 & 0.228 & H$_2$CO(3$_{12}$-2$_{11}$)&10.8 &  6430 & $-1.10\pm0.20$ & $+1.10\pm0.15$ & 1.1 & $2.0\times10^{-4}$  & 290$^3\rightarrow$1210 \\
01/16.56 & 0.228 & \multicolumn{8}{l}{CH$_3$OH($(4_0-4_{-1})+(5_0-5_{-1})+(6_0-6_{-1})+(7_0-7_{-1})$E)} \\
     & &                     & 15.5 &  9230 & $-1.45\pm0.15$ & $+1.90\pm0.30$ & 1.50 & $1.32\times10^{-5}$ &  5900--10600$^2$ \\
01/17.56 & 0.255 & HCN(3--2)     &  9.3 &  5620 & $-1.33\pm0.05$ & $+1.67\pm0.05$ & 1.45 & $1.60\times10^{-5}$ &  5890--12000$^2$ \\
01/17.55 & 0.255 & HNC(3--2)     &  9.2 &  5480 & $-1.23\pm0.11$ & $+1.60\pm0.15$ & 1.40 & $1.60\times10^{-5}$ &  5670--10500$^2$ \\
01/17.56 & 0.255 & CS(3--2)      & 17.0 & 10270 & $-1.17\pm0.05$ & $+1.20\pm0.10$ & 1.2  & 1--3$\times10^{-5}$& 2620--9800\\
01/17.59 & 0.256 & HCO+(1--0)    & 26.2 & 15830 & $-2.80\pm0.53$ & $+2.03\pm0.53$ & --   &  & \\
\hline
\end{tabular}
\end{center}
$^1$: Only low resolution (1~MHz = 1.2~\kms) spectra.\\
$^2$: Scale-length taking into account screening (Sect.4.2): based on the 
equivalent lifetime that would lead to the same number of molecule in the 
coma without screening.\\
$^3$: When we assume that H$_2$CO comes from an extended source with 
scale-length $1.75\times L_{\rm H_2CO}$, this value increases by a factor
$\approx4.23$ (to get a 1/e decrease from the peak value at 
$1.30\times L_{\rm H_2CO}$).
\end{table*}

\section{Production rates and abundances}
The production rates were computed using models that incorporate
collisions with neutrals and electrons, and radiative pumping
by the solar radiation \citep{Biv99,Biv00,Biv06a}. 
We did not consider infra-red pumping by the large 
and warm dust coma. Assuming a dust-to-gas ratio of 1.0 
(which might be underestimated for the 
dustier comets C/2006~P1 and C/2002~V1), and the simplified approach by 
\citet{Cro83}, we estimate that for HCN or CH$_3$OH in comet 
C/2006~P1 (McNaught), the vibrational pumping by the dust infrared radiation
is stronger than that by the solar radiation field within 2000--7000 km of the nucleus. 
However, collisions with water still dominate in a region twice as large
and control the excitation of the rotational levels.

We used the radial
profiles of the gas expansion velocity determined in the previous
section. In some cases, we considered different velocity profiles
to interpret the positive and negative sides of the lines, but
similar production rates were obtained as long as the velocity
profiles provided a good fit to the line widths. To compute
rotational level populations and collision rates with neutrals, we
used the gas temperatures given in Section 4.1. Inferred
production rates are given in Table~\ref{tabqp}. As discussed in
the next section, they are possibly wrong by a factor 1.5,
because of uncertainties in the model parameters.

\subsection{Uncertainties due to model parameters}
\label{sec:uncer}

The photodissociative lifetimes were taken from \citet{Cro94},
except for CS and HNC, which we constrained from the present
observations (Sect.~\ref{sec:cslife}). We took into account solar activity
when computing the photodissociation rates \citep{Cro89,Cro94,Boc85}.  
The uncertainty introduced by our simplifying assumptions in the
modeling of photolysis screening is not very large ($<15$\%, not
considering water production rate uncertainties).  Derived HCN
production rates are a factor between 1.4 and 2.5 smaller than
when this effect is omitted.

The main source of error in the computation of production rates
is the uncertainty in the gas kinetic temperature (except for
C/2002~X5 at $r_h>0.5$ AU, where the temperature is well
constrained). We assumed constant coma temperatures of 180, 150,
and 300~K for near perihelion data of comets C/2002~X5, C/2002~V1,
and C/2006~P1, respectively (Sect. 4.1). An increase (respectively
decrease) in these values by 50\% affects the production rate
determinations in the following way:
\begin{itemize}
\item $Q_{\rm HCN}$ and $Q_{\rm HNC}$ are increased by +33 to
+40\% (C/2006~P1) (respectively, decreased by 31\% to 39\%).

\item the same trend is observed for $Q_{\rm CS}$ and $Q_{\rm
CH_3CN}$: $\pm45$\% for a $\pm50$\% variation in the gas
temperature.

\item $Q_{\rm CH_3OH}$ is still more sensitive to the assumed
temperature, because of excitation of the torsional band at high $T$:
a $\pm50$\% change in temperature causes changes of +51/-39\%, +72/-53\%,
and +102/-64\%  in the production rates for C/2002~V1, C/2002~X5,
and C/2006~P1, respectively.

\item $Q_{\rm HC_3N}$ is minimal for $T\approx200$ K. The
retrieved $Q_{\rm HC_3N}$ is 50\% higher when either increasing or
decreasing the nominal temperature (180~K) by 50\%.

\end{itemize}

Therefore, most abundances relative to HCN are not significantly
sensitive to the assumed $T$. Two exceptions are the CH$_3$OH/HCN
(+20 to +40\% for $\Delta T$= +50\%) and HC$_3$N/HCN (+100\% for
$\Delta T$ = +50\%) ratios.

\subsection{Short- and long-term variations in the production rates}
\begin{figure}
\centering
\resizebox{\hsize}{!}{\includegraphics[angle=270]{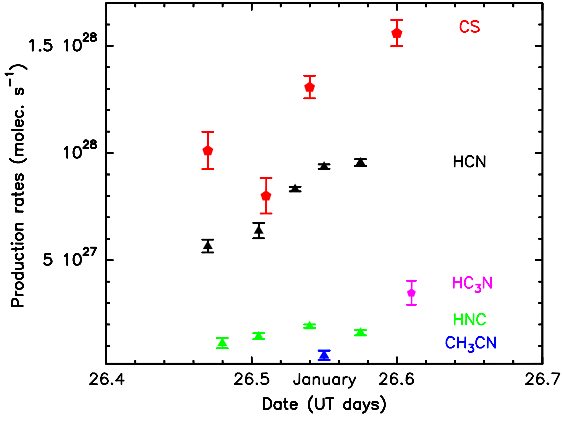}}
\caption{Molecular production rates in comet C/2002~X5 (Kudo-Fujikawa)
as a function of time on 26 January 2003: this plot shows the rapid increase
over $\approx$3~h of observations for HCN, CS, and HNC. The values for CH$_3$OH
and CH$_3$CN are the averages over the full period.}
\label{figqp02x5_peri}
\end{figure}

The most significant time variations are observed for C/2002~X5.
Rapid variations are observed around perihelion
(Fig.~\ref{figqp02x5_peri}). Strong variations in the light curve
and water production rate are also reported by \citet{Bou03} and
\citet{Pov03}. On 26.5 January 2003, the production rates of HCN,
CS and HNC increased by $\approx70$\% over a time interval of
$\approx$ 0.11 days (Table~\ref{tabqp}, Fig.~\ref{figqp02x5_peri}).

Over the 2--3 days of observations of comets C/2002~V1 and
C/2006~P1, production rates also varied. For C/2002~V1, an
heliocentric dependence $Q_{\rm CS}$ $\propto r_h^{-0.8}$ to
$r_h^{-1.5}$ is observed, while for C/2006~P1 $Q_{\rm HCN}\propto
r_h^{-4.5}$. For C/2006~P1, the variation in the HCN production rate
is steeper than the variation in the H$_2$O production rate
($2.4\times10^{29}r_h^{-3}$ \mols) deduced from the OH
observations (Table~\ref{tabqh2o}). This inconsistency could be
due to rotation-induced time variations and the flux loss 
cause by anomalous refraction on 17 January being 
unable to be quantified.

\begin{figure}
\centering
\resizebox{\hsize}{!}{\includegraphics[angle=0]{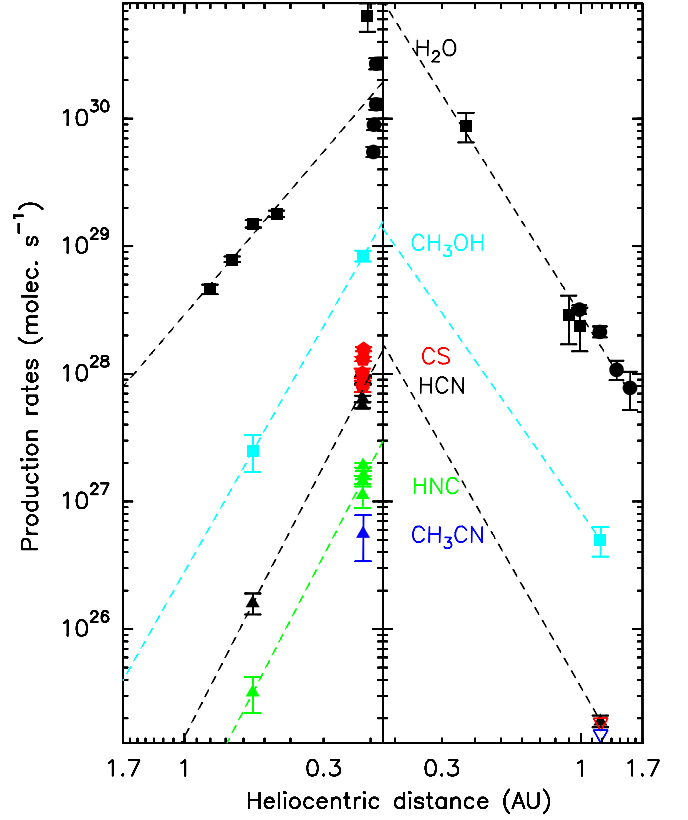}}
\caption{Molecular production rates in comet C/2002~X5 (Kudo-Fujikawa)
pre (left) and post (right) perihelion as a function of heliocentric
distance.
The water production rates
are based either on SOHO or Odin observations \citep{Pov03,Biv07a}
(black dots) or Nan\c{c}ay OH production rates multiplied by 1.1
(black squares). Dotted lines show the fitted evolutions (see text).}
\label{figqp02x5}
\end{figure}

The long-term (over 2 months) evolution of the production rates of
C/2002~X5 is shown in Fig.~\ref{figqp02x5}. The HCN, HNC, and CH$_3$OH
production rates vary according to $Q\propto r_h^{-3.5\pm0.5}$,
pre and post-perihelion, as for H$_2$O post-perihelion
($(2.6\pm0.2)\times10^{28}r_h^{-3.5\pm0.2}$ \mols).

\begin{table*}
\caption[]{Molecular production rates}\label{tabqp}
\begin{center}
\begin{tabular}{lcccccccccc}
\hline
UT date   &  $r_{h}$    & \multicolumn{9}{c}{Production rate in $10^{26}$\mols} \\[0cm]
{\small [mm/dd.d]} &    [AU]     & HCN       & HNC       & CH$_3$CN  & CH$_3$OH  & H$_2$CO$^1$ & CO  & CS     & HC$_3$N   & OCS \\[0cm]
\hline
\multicolumn{11}{l}{C/2002~X5 (Kudo-Fujikawa) (2003)}\\
\hline
01/13.61 & 0.553    &  $1.6\pm0.3$ & $0.32\pm0.10$ &               &   $25\pm8$  &         & $<95^2$ &              & & \\
01/26.47 & 0.214    & $56.6\pm3.0$ & $11.3\pm2.4$  &\multicolumn{1}{|l}{} & \multicolumn{1}{|l}{}  &    &         & $101.2\pm8.7$ & & \\
01/26.51 & 0.214    & $63.8\pm3.6$ & $14.5\pm1.4$  &\multicolumn{1}{|l}{} & \multicolumn{1}{|l}{}  &    &         &  $80.1\pm8.3$ & & \\
01/26.54 & 0.213    & $88.7\pm5.2$ & $19.2\pm0.8$  &\multicolumn{1}{|l}{$5.6\pm2.2$} & \multicolumn{1}{|l}{$841\pm82$}& &  & $130.8\pm5.3$ & & \\
01/26.58 & 0.212    & $95.5\pm1.6$ & $16.1\pm1.2$  &\multicolumn{1}{|l}{} & \multicolumn{1}{|l}{}  &    &         &             &  & \\
01/26.61 & 0.212    &              &               &\multicolumn{1}{|l}{} &\multicolumn{1}{|l}{}&  & & $156.0\pm6.1$ & $34.8\pm5.6$ & \\
03/12.72 & 1.184    & $0.19\pm0.02$&  $<0.064$     & $<0.15$       & $5.0\pm1.3$ &         &         & $<0.19$     &  & \\
\hline
\multicolumn{11}{l}{C/2002~V1 (NEAT) (2003)}\\
\hline
02/16.6  & 0.135    & $34.0\pm1.0$ & $10.0\pm0.9$  &\multicolumn{1}{|l}{$<5.6$} & $220\pm60$  &         &         & $102\pm 8$  &\multicolumn{1}{|l}{}        & $<460$ \\
02/17.5  & 0.108    & $41.1\pm2.5$ &               &\multicolumn{1}{|l}{}       & $261\pm50$  & $<389$  & $<1075$ & $144\pm27$  &\multicolumn{1}{|l}{$<10.3$} & $<581$ \\
\hline
\multicolumn{11}{l}{C/2006~P1 (McNaught) (2007)}\\
          &         & HCN          & HNC          & CH$_3$CN    & CH$_3$OH   & H$_2$CO$^1$ & CO       & CS     & HCOOH  & HDO \\[0cm]
\hline
01/15.65  & 0.207   &  $359\pm42$  &              &             &             &          & $<37500$   &        &       & \\
01/16.54  & 0.228   &  $286\pm28$  &              &             &\multicolumn{1}{|c}{$1450$}  & $590$    & & & $222\pm60^3$ & 90  \\
01/16.58  & 0.229   &  $242\pm23$  &              &             &\multicolumn{1}{|c}{$\pm100$}& $\pm110$ & &   &       & $\pm30^3$\\
01/17.55  & 0.255   &  $146\pm14$  & $16.7\pm2.7$ & $10.6\pm4$  &  $645\pm77$ &          &             & $217\pm33$ &  & \\
01/17.58  & 0.256   &  $142\pm15$  &              &             &             &          & $4290^3$    &       &       & \\
          &         &              &              &             &             &          & $\pm1380$   &       &       & \\
\hline
\end{tabular}
\end{center}
$^1$: Assuming that H$_2$CO comes from an extended source with scale-length
$1.75\times L_{\rm H_2CO}$.\\
$^2$: Uncertainty in the tuning: the backend may not have been properly connected to the receiver.\\
$^3$: Marginal values.
\end{table*}

\subsection{Molecular abundances and discussion}
    Table~\ref{tababund} summarizes the molecular abundances relative
to water and HCN in the three comets.

The water production rates used as reference for the short
heliocentric distances ($r_h<0.3$ AU) are uncertain, as discussed
in Section 3. For C/2002~X5, according to the short-term
variations seen for all molecules, we assumed $Q_{\rm H_2O} =
35+10\times\sin{\left(2\pi\frac{t[d]-26.54}{0.2}\right)}\times10^{29}$ \mols.
For C/2006~P1, we used $Q_{\rm H_2O} = 2.4\times10^{29}r_h^{-3}$
\mols, and for C/2002~V1 $Q_{\rm H_2O} = 2.7\times10^{29}r_h^{-1}$
\mols, these values being possibly in error by $\pm50$\%.
Resulting abundances relative to water are roughly in agreement
with abundances measured in other comets \citep{Biv02a}.
The CH$_3$OH/H$_2$O ratio seems however quite low in comet C/2002~V1 and
especially in C/2006~P1. The HDO/H$_2$O ratio in comet
C/2006~P1, is not well constrained, both OH and HDO
being only marginally detected, but production rates are
compatible with the ratio measured in other comets 
\citep[$\approx6\times10^{-4}$]{Boc98,Mei98,Jeh09}.

We now turn our discussion to abundances relative to HCN,
which are more reliable since HCN was always observed
simultaneously. The main observed features are:

\begin{itemize}
\item A decrease in the CH$_3$OH/HCN ratio at low $r_h$ $ \propto
r_h^{0.5}$ is observed for C/2002~X5, confirming the trend
observed in comet C/1996~B2 (Hyakutake) \citep{Biv99}. Using this
$r_h^{0.5}$ variation, we infer CH$_3$OH/HCN ratios at 1 AU of 18
and 10 for C/2002~V1 and C/2006~P1, respectively, which are within
the range of abundances measured in comets, though C/2006~P1
belongs to the methanol-poor category \citep{Biv02a}.

\item The CS/HCN ratio varies as $0.35\times r_h^{-1.0}$. A similar
heliocentric dependence ($r_h^{-0.8}$) was measured in comet
C/1996~B2 \citep{Biv99}. In comet 153P, CS/HCN followed
$0.55r_h^{-0.7}$ \citep{Biv06a}. The puzzling heliocentric
variation in the CS/HCN ratio is confirmed down to very low $r_h$.

\item Measuring the HNC/HCN ratio at small $r_h$ was one of the
main objectives of our analysis of these observations. HNC was easily detected in
the three comets with an abundance ratio HNC/HCN between 0.11 and
0.29 at $r_h < 0.26$ AU. The measurements in comets C/2002~X5 and
C/2002~V1 were included in a previous study of the heliocentric 
variation of HNC/HCN based on a sample of 11 comets at $r_h$ in the range
0.14--1.5 AU \citep{Lis08}. The HNC/HCN sample was fitted by a
power law in $r_h^{-2.3}$, with an indication being found of a possible
flattening at $r_h$ $<$ 0.5 AU. The values inferred for our three
comets do not show a trend for the more productive comets being
enriched in HNC, as expected for a formation of HNC by chemical
reactions \citep{Rod98}, but rather the opposite. The HNC/HCN
ratio in C/2006~P1 is 0.13, smaller than the value in comet
Hale-Bopp at 1 AU ($\approx0.25$), this comet being slightly
more active than comet Hale-Bopp. Possible origins of HNC in
cometary atmospheres are discussed in \citet{Rod98} and
\citet{Lis08}. Destruction of HNC by reaction processes possibly
takes place at low $r_h$. Small differences in the line shapes between
HCN and HNC suggest that HNC might be partly produced in the coma
and destroyed further away from the nucleus or be more sensitive
to photodissociation than HCN. 
\end{itemize}

Comet C/2006~P1 has compositional similarities with fragments B and C
of comet 73P/Schwassmann-Wachmann 3. All are CH$_3$OH and CO-poor, 
and the H$_2$CO/HCN,
CH$_3$CN/HCN ratios are similar \citep{Biv06b,Del07b}. 
However, C/2006~P1 is rich in volatile hydrocarbons and in
NH$_3$, while 73P is depleted in these compounds according 
to \citet{Del07a}.

Comet C/2002~X5, on the other hand, has a normal CH$_3$OH
abundance. It is unusually rich in HC$_3$N. The HC$_3$N/HCN ratio
is four times higher than in comet Hale-Bopp \citep{Boc00}, and
still higher than upper limits found in some comets
\citep{Biv06a}. Interestingly, C/2002~X5 belongs to the class of
carbon-rich comets \citep{Pov03}. Finally, the large abundances
of HC$_3$N, HNC, and CH$_3$CN relative to HCN in this comet imply
that these species are significant contributors to the production
of CN radicals in this comet.

\begin{table*}
\caption[]{Relative abundances}\label{tababund}
\begin{center}
\begin{tabular}{lrrrr}
\hline
Molecule & \multicolumn{2}{r}{C/2002~X5 (Kudo-Fujikawa)} & C/2002~V1 (NEAT) & C/2006~P1 (McNaught) \\
\hline
$<r_h>^1$& 0.5--1.2              &   0.21           &       0.12     &       0.24      \\
\hline
\multicolumn{5}{l}{Abundance relative to water}\\
\hline
HCN      & $0.11\pm0.02$\% & $0.23\pm0.02$\%   & $0.17\pm0.01$ \%  & $0.13\pm0.02$ \%     \\
HNC     &$0.027$--$<0.03$\%& $0.047\pm0.008$\% & $0.05\pm0.005$\%  & $0.011\pm0.002$\% \\
CH$_3$CN &                 & $0.016\pm0.006$\% & $<0.024$ \%       & $0.008\pm0.003$\% \\
HC$_3$N  &                 & $0.081\pm0.013$\% & $<0.045$ \%       & \\
CH$_3$OH & $2.3\pm0.5$ \%  & $2.4\pm0.2$\%     & $1.1\pm0.2$ \%    & $0.6\pm0.2$\% \\
H$_2$CO  &                 &                   & $<1.95$ \%        & $0.29\pm0.06$\% \\
CO       & $<8$ \%         &                   & $<4.3$ \%          & $3\pm1$\% \\
HCOOH    &                 &  $<0.9$ \%        &                   & $0.11\pm0.03$\% \\
HDO      &                 &                   &                   & $0.045\pm0.015$\% \\
CS       & $<0.088$\%      & $0.36\pm0.02$\%   & $0.54\pm0.08$\%   & $0.15\pm0.02$\% \\
OCS      &                 &                   & $<1.54$ \%        & \\
SO       &                 &                   & $<0.81$ \%        & \\
SiO      &                 & $<0.008$ \%       & $<0.011$\%        & \\
\hline
\multicolumn{5}{l}{Abundance relative to HCN}\\
\hline
HNC &$0.20\pm0.07$--$<0.33$& $0.198\pm0.024$   & $0.29\pm0.028$    & $0.114\pm0.026$ \\
CH$_3$CN &                 & $0.07\pm0.03$     &   $<0.14$         & $0.08\pm0.03$ \\
HC$_3$N  &                 & $0.36\pm0.06$     &   $<0.25$         & \\
CH$_3$OH & $20\pm5$        & $11\pm1$          & $6.4\pm1.0$       & $5.1\pm0.3$ \\
H$_2$CO  &                 &                   & $<9.5$            & $2.2\pm0.4$ \\
CO       & $<60$           &                   & $<26$             & $31\pm10$ \\
HCOOH    &                 &  $<4$             &                   & $0.8\pm0.2$ \\
HDO      &                 &                   &                   & $0.3\pm0.1$ \\
CS       & $<0.96$         & $1.55\pm0.17$     & $3.2\pm0.3$       & $1.51\pm0.24$ \\
OCS      &                 &                   & $<8.6$            & \\
SiO      &                 & $<0.04$           & $<0.068$          & \\
\hline
\end{tabular}
\end{center}
\end{table*}

\subsection{Search for refractory molecules}
    SiO was searched through its $J$(6--5) transition at 260518.020 MHz
 in comets C/2002~X5 and C/2002~V1 around perihelion
(Tables~\ref{tabobs02x5},\ref{tabobs02v1}). It was not detected
down to about 0.01\% relative to water, assuming thermal
equilibrium and a photodissociation rate of $10^{-5}$ s$^{-1}$.
The same tuning at 260~GHz also allowed us to search for NaCl 
$J$(20--19) line in C/2002~V1, while the KCl $J$(30--29) line 
was searched for at 230~GHz in C/2002~V1 and C/2006~P1, in both 
case without success.

\section{Conclusion}

We have reported on our analysis of millimeter spectroscopic observations 
of three bright comets
(C/2002~X5 (Kudo-Fujikawa), C/2002~V1 (NEAT), and C/2006~P1
(McNaught)) when they were close to the Sun ($r_h <$ 0.25~AU).
Our main challenge has been to cope with
ephemeris uncertainties by searching for the peak of brightness in
HCN emission using coarse mapping. 
We have obtained the following results:
\begin{itemize}

\item The screening effect in photolytic processes, which affects the 
distribution of the molecules in the coma, was modeled to determine 
accurate production rates. This effect is significant, especially
for C/2006~P1 where it increases by 40 to 60\% the number of
molecules in the coma. 

\item The various lines have different widths. Velocity variations in the
coma were considered to interpret the line shapes. This 
provided constraints on the CS and HNC scale-lengths. 

\item The CS photodissociation rate was estimated to
$2.5\pm0.5\times10^{-5}$ s$^{-1}$ ($r_h$ = 1 AU), which is
compatible with other estimates \citep{Boi07}.

\item The HNC photodissociation lifetime is found to be slightly
shorter (by $\approx30$\%) than the lifetime of HCN, though this
estimate does not consider destruction paths of HNC by chemical
reactions.

\item The CS/HCN production rate ratio in cometary atmospheres
follows a  heliocentric dependence $\approx r_h^{-0.8}$ down to
$r_h = 0.1$ AU. We rule out artefacts related to uncertainties in
CS lifetime. The origin of this variation remains unexplained.

\item The CH$_3$OH/HCN production rate ratio decreases with
decreasing $r_h$ according to $r_h^{0.5}$.

\item HNC/HCN ratios are high (0.11 to 0.29) in the range $r_h$ =
0.14--0.25 AU, consistent with HNC being a by-product of the
thermal degradation of organic grains \citep{Lis08}. A lower HNC
abundance is measured for the more productive comet, possibly
indicating destruction of this reactive molecule by chemical
processes.

\item The H$_2$CO/HCN abundance ratio measured in C/2006~P1 (2.2 at 0.23
AU) and C/2002~V1 ($<9.5$ at 0.12 AU) is in the range of values
measured in comets at 1 AU from the Sun \citep[1.6--10]{Biv02a,Biv06a}. 
This possibly rules out H$_2$CO production from the
thermal degradation of polymers as proposed by \citet{Fra06}.

\item The HC$_3$N abundance in C/2002~X5 is higher, by a factor of
four or more, than in any of the other eight comets in which it had
previously been measured. HC$_3$N, CH$_3$CN, and HNC are all significant
contributors to the production of CN radicals in this comet.

\item Searches for SiO, NaCl, and KCl were unsuccessful. The upper
limit set for SiO is $<10^{-4}$ relative to H$_2$O.

\end{itemize}

\begin{acknowledgements}
        We are grateful to the IRAM staff and
to other observers for their assistance during the observations.
IRAM is an international institute co-funded by the
Centre National de la recherche scientifique (CNRS), the Max Planck
Gesellschaft and the Instituto Geogr\'afico Nacional, Spain.
This research has been supported by the CNRS and the
Programme national de plan\'etologie de l'Institut des sciences de l'univers
(INSU).
\end{acknowledgements}

\end{document}